\documentclass[11pt,letterpaper]{article}
\usepackage{arxiv}

\usepackage{graphicx}
\usepackage{rotate}
\usepackage{wrapfig}
\usepackage{caption}
\usepackage[T1]{fontenc}
\usepackage[english]{babel}
\usepackage{times}
\usepackage{amsmath,amstext,amsbsy,amsopn,amsthm,amsfonts,amssymb}
\usepackage{mathtools}
\usepackage{url}
\usepackage{tabularx,booktabs}
\newcolumntype{C}{>{\centering\arraybackslash}X}
\usepackage{booktabs,array,colortbl}
\usepackage{lastpage}
\usepackage{multirow}
\usepackage{algorithm}
\usepackage{balance}
\usepackage{placeins}
\usepackage{array}
\usepackage{blindtext}
\usepackage{subfigure}
\usepackage{caption}
\usepackage{capt-of}
\usepackage{booktabs}
\usepackage[small]{titlesec}
\usepackage{paralist}
\usepackage{verbatim} 
\usepackage[usenames,dvipsnames,svgnames]{xcolor}
\usepackage{algorithmic}
\usepackage{fancyhdr}
\usepackage{wrapfig}
\providecommand{\keywords}[1]{\textbf{\textit{Keywords---}} #1}
\begin{document}

\title{iGLU 1.0: An Accurate Non-Invasive Near-Infrared
	Dual Short Wavelengths Spectroscopy based
	Glucometer for Smart Healthcare}



\author{
\begin{tabular}{ccc}
Prateek Jain & Amit M. Joshi & Saraju P. Mohanty \\
Electronics \& Commu. Eng. & Electronics \& Commu. Eng. & Computer Science and Engineering   \\
MNIT, Jaipur, India. & 	MNIT, Jaipur, India. & University of North Texas, USA. \\
prtk.ieju@gmail.com &amjoshi.ece@mnit.ac.in & saraju.mohanty@unt.edu
\end{tabular}	
}

\maketitle

\cfoot{Page -- \thepage-of-\pageref{LastPage}}

\begin{abstract}
In the case of diabetes, fingertip pricking for a blood sample is inconvenient for glucose measurement. Invasive approaches like laboratory test and one-touch glucometer enhance the risk of blood-related infections. To mitigate this important issue, in the current paper, we propose a novel Internet-of-Medical-Things (IoMT) enabled edge-device for precise, non-invasive blood glucose measurement. In this work, a near-infrared (NIR) spectroscopic technique using two wavelengths (940 nm, 1300 nm) is taken to detect the glucose molecule from human blood. The novel device called iGLU is based on NIR spectroscopy and machine learning (ML) models of high accuracy. An optimal multiple polynomial regression model and deep neural network (DNN) model have been presented for precise measurement. The proposed device is validated and blood glucose values are stored on the cloud using open IoT platform for remote monitoring by an endocrinologist. For device validation, the estimated blood glucose values have been compared with blood glucose values obtained from the invasive device. It has been observed that mean absolute relative difference (MARD) and average error (AvgE) are found 4.66\% and 4.61\% respectively from predicted blood glucose concentration values. The regression coefficient is found 0.81. The proposed spectroscopic non-invasive device provides accurate and cost-effective solution for smart healthcare. 
\end{abstract}



\section{Introduction}

The ubiquity of diabetic patients has become double from 2010 over the world \cite{habbu2019estimation}. The estimated diabetes dissemination from 2009 is 290 million and is expected to affect 450 million people by 2030. Diabetes takes place when a person faced difficulty in balancing the body glucose level in different prandial states \cite{1186525}. Diabetes is caused by the deficiency of insulin with respect to the generated glucose in the body. It may be due to demolition of insulin which is produced by beta cells in pancreas \cite{6778812}. Diabetes may also be caused by insulin resistance. This is a condition in which the muscles, fat and liver cells of the body do not consume insulin effectively \cite{5291722}. Diabetes is classified into three parts: Type 1 diabetes, Type 2 diabetes and gestational diabetes \cite{7576627}. In type 1 diabetes, the immune system of the body attacks and destroys the cells of the pancreas which produce insulin \cite{7933990}. This results in the affected person who will be unable to generate insulin naturally. Type 2 diabetes is the most common diabetic stage which is most commonly seen in the people over the world. In this type of diabetes, the pancreas will be able to generate some amount of insulin. Gestational diabetes occurs in women in the later stages of pregnancy. Most common symptoms of diabetes are the excretion of urine within short durations, consistently hungriness, thirsty, unexpected weight loss, tiredness and vision changes \cite{1186525}. The long duration of diabetes without any treatment may cause kidney disease, stroke, heart disease, nerve damages and blindness. After being these problems, probability of death with diabetes has become 50\% higher than without diabetes in adults \cite{Dhar2013}. Diabetes may be controlled through physical exercise, diet, and proper use of insulin regimen. Oral medications are also useful to control for an early stage of diabetes. Controlling of diabetes also includes reduction of risk factors for cardiovascular disease such as lipid profile, high systolic and diastolic blood pressure. In most cases of adults, 5 \% Type 1 diabetic patients have been considered approximately in all diagnosed case. Whereas, 90-95 \% Type 2 diabetic patients have been considered for treatment. Hence, it is essential to develop the device of blood glucose measurement for rapid and frequent diagnosis of diabetes \cite{4956982}. People will be more conscious of controlling blood glucose with frequent monitoring. Conventional blood glucose detection method for diabetic patients is a chemical process using a drop of blood. A diabetic patient needs to measure blood glucose frequently \cite{1186525}. Interaction of lancets with blood may increase the chance of blood-related diseases and trauma. Frequent use of invasive device enhances the probability of blood infection through lancets \cite{Reddy2017}. Hence, such type of invasive method for glucose measurement is not advisable in case of frequent blood glucose monitoring. Therefore, it became important to design the non-invasive device for clinical tests, which is beneficial for health care. Designing a non-invasive blood glucose monitoring device involves spectroscopy techniques. Spectroscopy is the interaction between matter and optical radiation.

\begin{figure}[htbp]
	\centering
	\includegraphics[width=0.85\textwidth]{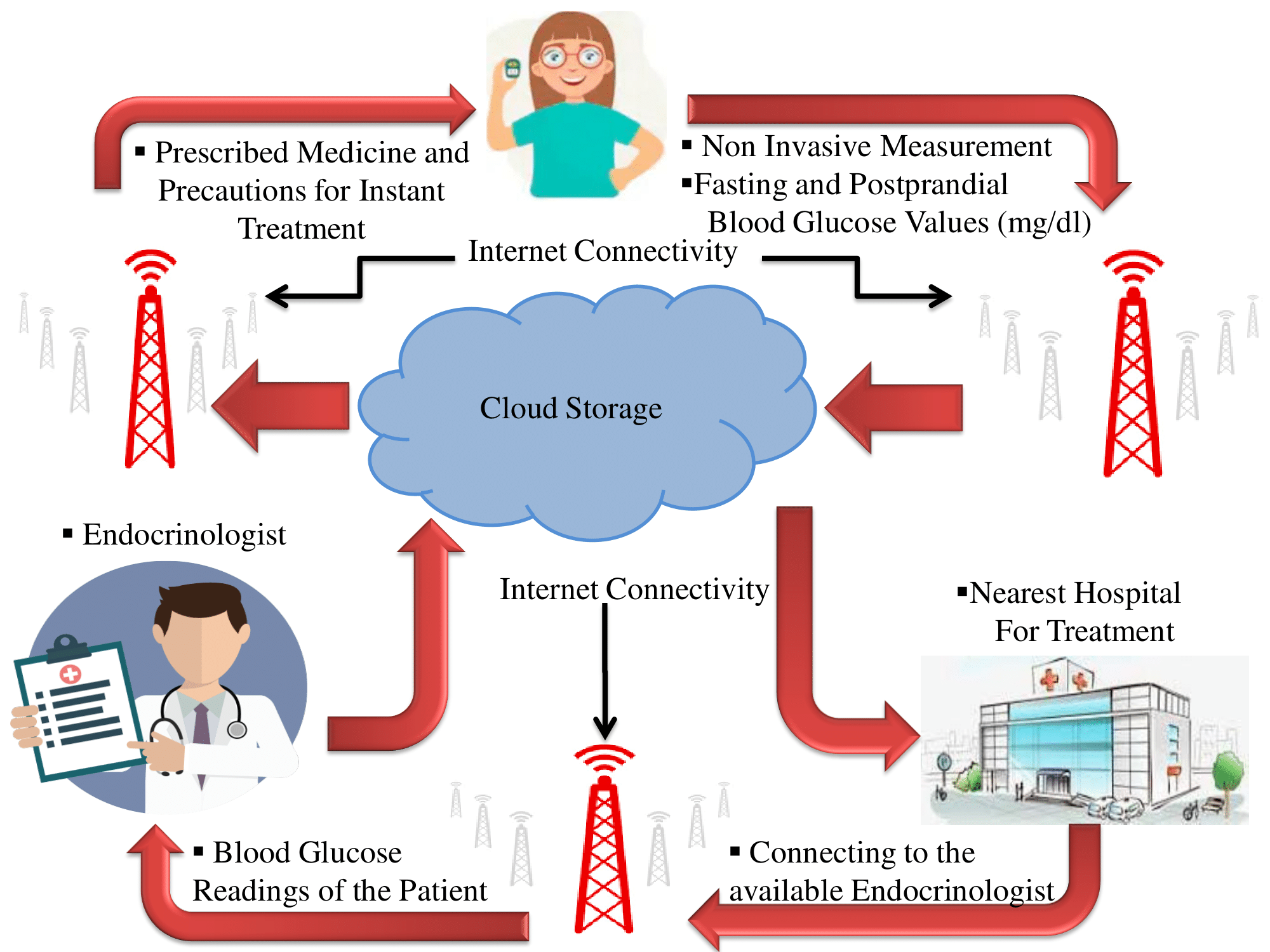}
	\caption{Blood glucose diagnosis in smart health care system \cite{Jain_IEEE-MCE_2020-Jan_iGLU1}.}
	\label{smart}
\end{figure}

The effective and affordable solutions are the key enabler of smart healthcare in the big picture of smart cities \cite{Mohanty_CEM_2016-July}. These are being envisioned to sustain the population migration to urban areas \cite{Mohanty_CEM_2016-July}. Smart healthcare system comprises of ambient intelligence, quality of service and also offers continuous support of the monitoring of critical diseases \cite{Sundaravadivel_CEM_2018-Jan, Sundaravadivel_IEEE-TCE_2018-Aug}. The smart healthcare system is most demandable for remote monitoring of diabetic patients with low cost and rapid diagnosis \cite{Mohanty_POT_2006-Mar}. Traditional blood glucose measurement is unable to serve everyone's need in a rural and remote location. Despite having good diagnostic centres for clinical test facility in the urban area, medical services are not approachable to everyone at a remote location. It is necessary to monitor blood glucose of diabetic patients where the diagnosis facility is not easily available. The instant diagnoses of blood glucose and frequent monitoring are the recent challenges in the smart healthcare system. The process flow of blood glucose diagnosis in the smart health care system is shown in Fig. \ref{smart}.

IoMT is considered as a specific Internet-of-Things (IoT) paradigm in the case of smart healthcare with medical-Things \cite{Rachakonda_ICCE_2019}. IoMT enabled handheld non-invasive glucose measurement end-device has a strong potential for rapid monitoring as well as to facilitate the interaction with an endocrinologist to the remote located diabetic patients where diagnosis centres and hospitals are not easily available. According to this environment, patients measure their blood glucose without pricking blood and directly store to the cloud where nearby endocrinologist can monitor the glucose data of each patient. The prescription would also be provided by an endocrinologist to the remotely located patient for further treatment. 

\begin{figure}[htbp]
	\centering
	\includegraphics[width=0.85\textwidth]{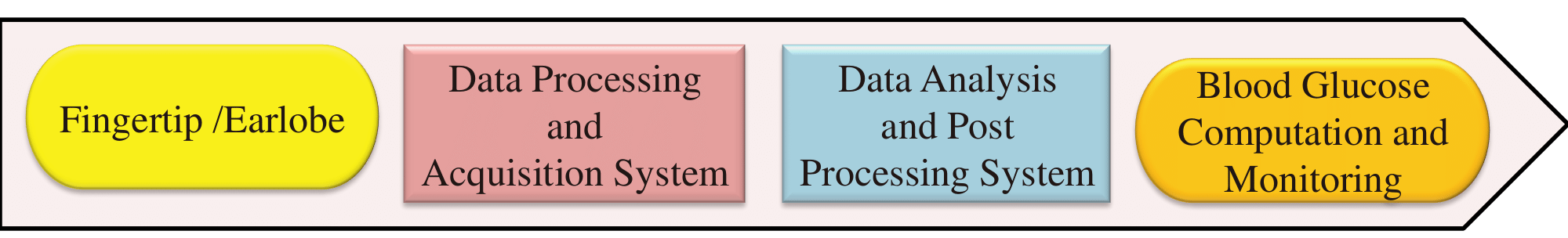}
	\caption{Processing steps of non invasive blood glucose measurement device.}
	\label{theme}
\end{figure}

This technique is used to measure the light intensity after interaction with matter at specific wavelengths. In the proposed non-invasive measurement, optical detection is involved. The collected data is processed for acquisition and blood glucose is estimated by machine learning-based computation model and the whole process flow is represented in Fig. \ref{theme}. Till date, several devices have been designed and developed for non-invasive blood glucose measurement. Some products such as glucotrack, glucowise, DiaMon Tech and device from CNOGA medical etc. are still not commercialized. However, they have limitations in terms of precise measurement. The cost of the product is also high which varies in the range of 300-400 USD. Therefore, the cost-effective solution for non-invasive blood glucose measurement is addressed in this paper.

The rest of the paper is organized in the following manner. The related prior works are described in Section \ref{Sec:Prior-Works}. The novel contribution of current work is represented in Section \ref{Sec:Novel}. The proposed device is elaborated with the mechanism of glucose detection in Section \ref{Sec:Proposed-Work}. Section \ref{Sec:Calibration} discusses analytical modelling and calibration details of the sensor. Experiments, analytical modelling and error analysis have been presented in Section \ref{Sec:Experimental-Results}.

\section{State-of-Art in Blood Glucose-Level Measurement}
\label{Sec:Prior-Works}

\begin{figure}[htbp]
	\centering
	\includegraphics[width=0.999\textwidth]{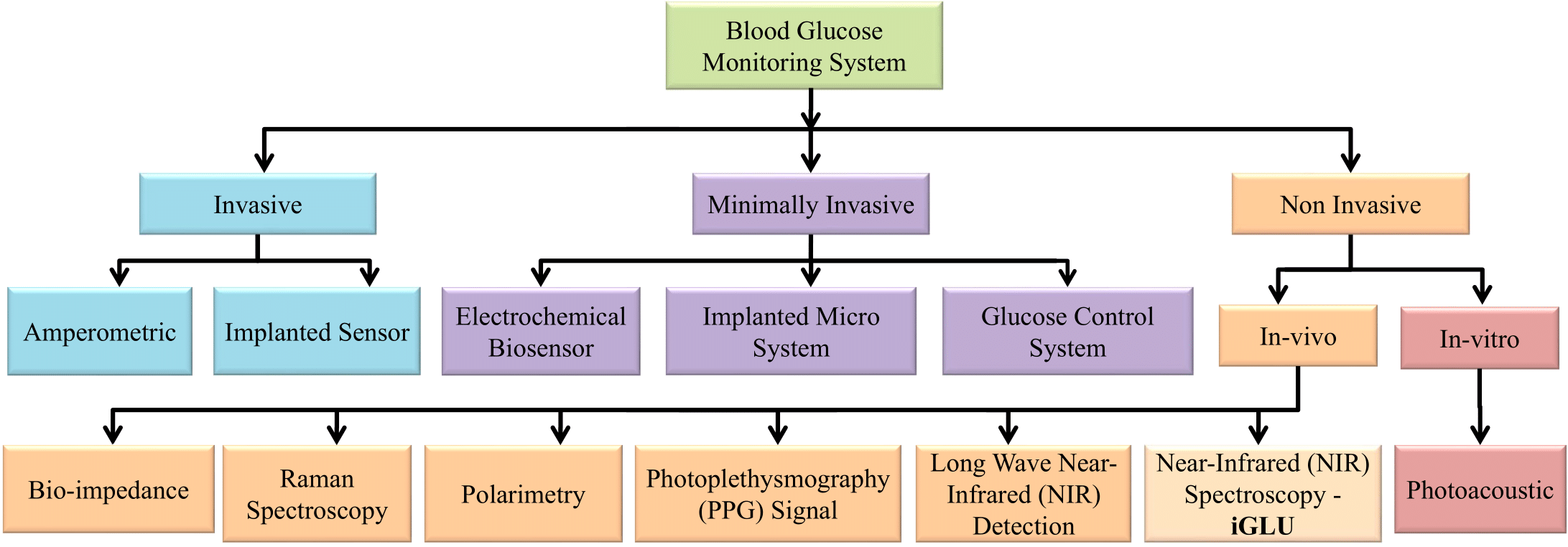}
	\caption{An overview of various blood glucose-level measurement devices or systems \cite{Jain_IEEE-MCE_2020-Jan_iGLU1}.}
	\label{priorwork}
\end{figure}

Blood glucose measurement is possible using invasive, minimally invasive and non-invasive methods (Figure \ref{priorwork}). Frequent pricking, as needed in invasive methods, for glucose measurement causes trauma. Therefore, the semi-invasive approach has the advantage of continuous glucose monitoring without multiple times pricking. However, non-invasive methods can completely eliminate pricking which opens door to painless and continuous glucose monitoring (CGM).

\begin{figure}[htbp]
	\centering
	\includegraphics[width=0.90\textwidth]{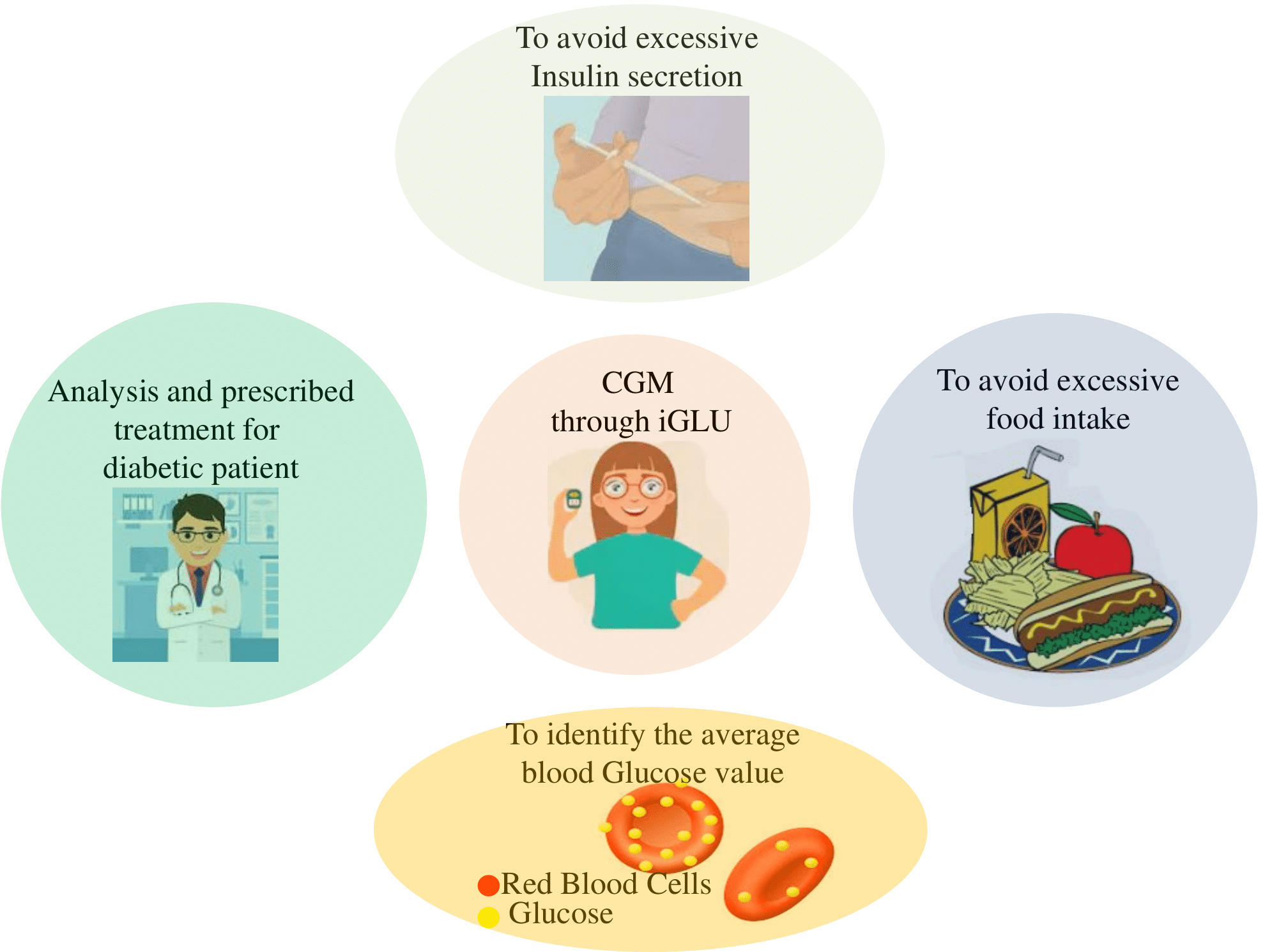}
	\caption{The objectives of continuous glucose monitoring}
	\label{idea}
\end{figure}

Continuous glucose monitoring is necessary to analyze the blood glucose level in each prandial state. It helps to control the glucose level after taking medicines, physical activity and insulin secretion. Diabetologist can monitor glucose level and suggest the prescribed treatment according to the condition. CGM is also useful for type 1 diabetic patients. This may reduce the chance to take excessive insulin secretion. During CGM, the diet can be controlled with frequent glucose monitoring. The motive of CGM is represented in Fig. \ref{idea}.

Continuous glucose monitoring helps the patient to measure glucose frequently. Because of this, an average glucose value can be identified for a long time. This is helpful to measure the glucose values for the last 90 days, by which glycated haemoglobin (HbA1c) can be determined.

\subsection{Invasive Methods}
A low-invasive amperometric glucose monitoring biosensor has been proposed using fine pointed glucose oxidase immobilized electrode which doesn't require more than 1mm in length to be inserted in skin \cite{li2016fine}. 
The photometric approach has been explored for glucose measurement using small blood volumes \cite{demitri2017measuring}. The issue of high volume of pricking blood has been solved by this system for testing. 
A fully implanted first-generation prototype sensor has been presented for long-term monitoring of subcutaneous tissue glucose \cite{lucisano2017glucose}. This wearable sensor which is integrated as an implant is based on a membrane containing immobilized glucose oxidase and catalase coupled to oxygen electrodes, and a telemetry system. 

\subsection{Minimally Invasive Methods}
Implantable biosensors have been deployed for continuous glucose monitoring \cite{7576627}. Wearable minimally invasive microsystem has been explored for glucose monitoring \cite{7933990}. A microsystem has been presented for glucose monitoring which consists of microfabricated biosensor flip-chip bonded to a transponder chip \cite{4956982}. The output signal has been measured by this transponder chip of the biosensor and transmitted the measured data back to the external reader. A method has been discussed to reduce the frequency of calibration of minimally invasive Dexcom sensor \cite{acciaroli2018reduction}. An artificial pancreas has been represented along with a glucose sensor to control diabetes \cite{6778812}. But, approaches based semi-invasive devices have not been tried for real-time application. These wearable microsystems are neither painless nor cost-effective solutions.

\subsection{Non-invasive Methods}
To make the painless system, photoacoustic spectroscopy has been introduced for non-invasive glucose measurement \cite{Pai2018}. However, utilization of the LASER makes the setup costly and bulky. Hence, it would not provide a viable solution for commercialization. 
An enzyme sensor has been explored for glucose measurement in saliva \cite{677170}. Glucose detection is possible using Intensity Modulated Photocurrent Spectroscopy (IMPS) spectroscopy that connects the electrodes to the skin which is affected by sweat \cite{Song2015}. The high precision level is possible through these methods as sweat and saliva properties vary for individuals. Hence, this approach is not suitable for glucose measurement. 
The blood glucose measurement has also been explored using Raman spectroscopy in laboratory \cite{shih2015noninvasive}. The experimental setup for Raman spectroscopy required a large area and will not be portable. Glucose measurement has also been done from the anterior chamber of the eye which limits its usage of continuous monitoring \cite{pirnstill2012vivo}. Blood glucose has been estimated using photoplethysmography (PPG) signal \cite{monte2011non, habbu2019estimation}. 
The functional relationship has been derived from PPG signal for glucose estimation \cite{Tsai_MCE-2020-Jan_PPG-Glucose}. Different features have been extracted from logged PPG signal for glucose estimation \cite{habbu2019estimation}.

\subsection{Consumer Electronics for Glucose-Level Monitoring}
Several devices have been developed for non-invasive blood glucose measurement. Some products such as glucotrack, glutrac, glucowise, DiaMon Tech and device from CNOGA medical are not commercialized. Glutrac is multi-parameter health test device with smart healthcare. However, they have limitations in terms of precise measurement. The cost of the product is also high which varies in the range of 300-400 USD. Therefore, the cost-effective solution for non-invasive blood glucose measurement is needed. In this way, SugarBEAT from Nemaura medical, Freestyle Libre sensor are launched in the market for medication. These are adhesive and daily disposable skin-patch for continuous monitoring. These are not completely non-invasive device for glucose monitoring. Hence, these may not be advisable as a non-invasive continuous glucose monitoring device. Omelon B-2 claims that the device is non-invasive stripless for continuous glucose monitoring. The device is launched for glucose level monitoring. But, precise measurement has not been possible as per the public remarks. Prototypes of the Glucosense Monitor which is based on fluorescent technique, have been made, but a lot of work has to be done for commercialization. The researchers from the university of texas of Dallas developed a flexible textile-based biosensor for glucose level investigation. Still, the sensing device has limitations in terms of precise glucose measurement. 

\subsection{Wearable versus Non-wearable for Glucose Monitoring}
Prior works related to glucose monitoring have been discussed which represent wearable and non-wearable approaches. Raman spectroscopy, photoacoustic spectroscopy and invasive approach based systems are not wearable. Minimally invasive devices which have been discussed, are implantable. Minimally invasive biosensors such as SugarBEAT from Nemaura medical, Freestyle Libre sensor and Dexcom sensor are wearable sensors. Glucowatch is a wearable device. Smart contact lenses and sweat patches are wearable devices. LifePlus has launched the first non-invasive continuous glucose monitoring (CGM) wearable device. However, the company did not give the details of the technical specifications. Now, the device is under trials for commercialization. Other approaches based non-invasive device are wearable. Here, iGLU is non-invasive, optical detection based wearable device for continuous glucose monitoring with IoMT framework.

\section{Novel Device iGLU to Advance the State-of-Art in Wearable for Continuous Blood Glucose Monitoring}
\label{Sec:Novel}

Non-invasive detection overcomes the chance of being blood-related diseases. However, this approach has some limitations such as large set-up, measuring object (retina) and skin properties (including dielectric constant and sweat level). Therefore, a portable non-invasive precise glucose measurement device for continuous monitoring is needed. An initial example is a non-invasive glucose measurement using NIR spectroscopy and Huber's regression model \cite{jain2019precise}. 
The wavelengths are judiciously selected after experimental analysis which has been done in material research center MNIT, Jaipur (India). The reported system was calibrated with healthy samples and validated through one diabetic sample. The smart healthcare in IoMT framework is also needed to provide remote access through blood glucose monitoring. Hence, current research is presented using an optimized and proposed regression model with IoMT framework. 

\subsection{Research Question and Challenges Addressed in the current Paper}

There are several glucose monitoring systems which neither provide precise measurement nor a cost-effective solution. These systems are not enabled for smart healthcare. The following questions are resolved in iGLU for the advancement of smart healthcare: 
\begin{enumerate}
	\item 
	How can we have a device that automatically
	performs all the tasks of blood-glucose monitoring at
	the user and stores the data in cloud for future use by
	the patient and healthcare providers? 
	
	\item 
	Can we have a
	the device that automatically performs all the tasks of blood glucose
	at the user-end so that it can function even if the
	Internet connectivity is not available all the time? 
	
	\item 
	Can we have a device that can perform automatically blood glucose
	non-invasive to avoid hassle and risky finger pricking all the time monitoring is needed? 
	
	\item 
	Can we have a device that can measure the blood glucose of all
	types of patients precisely?
\end{enumerate}

\subsection{Novel Contributions of the current Paper}

\begin{figure}[htbp]
	\centering
	\includegraphics[width=0.85\textwidth]{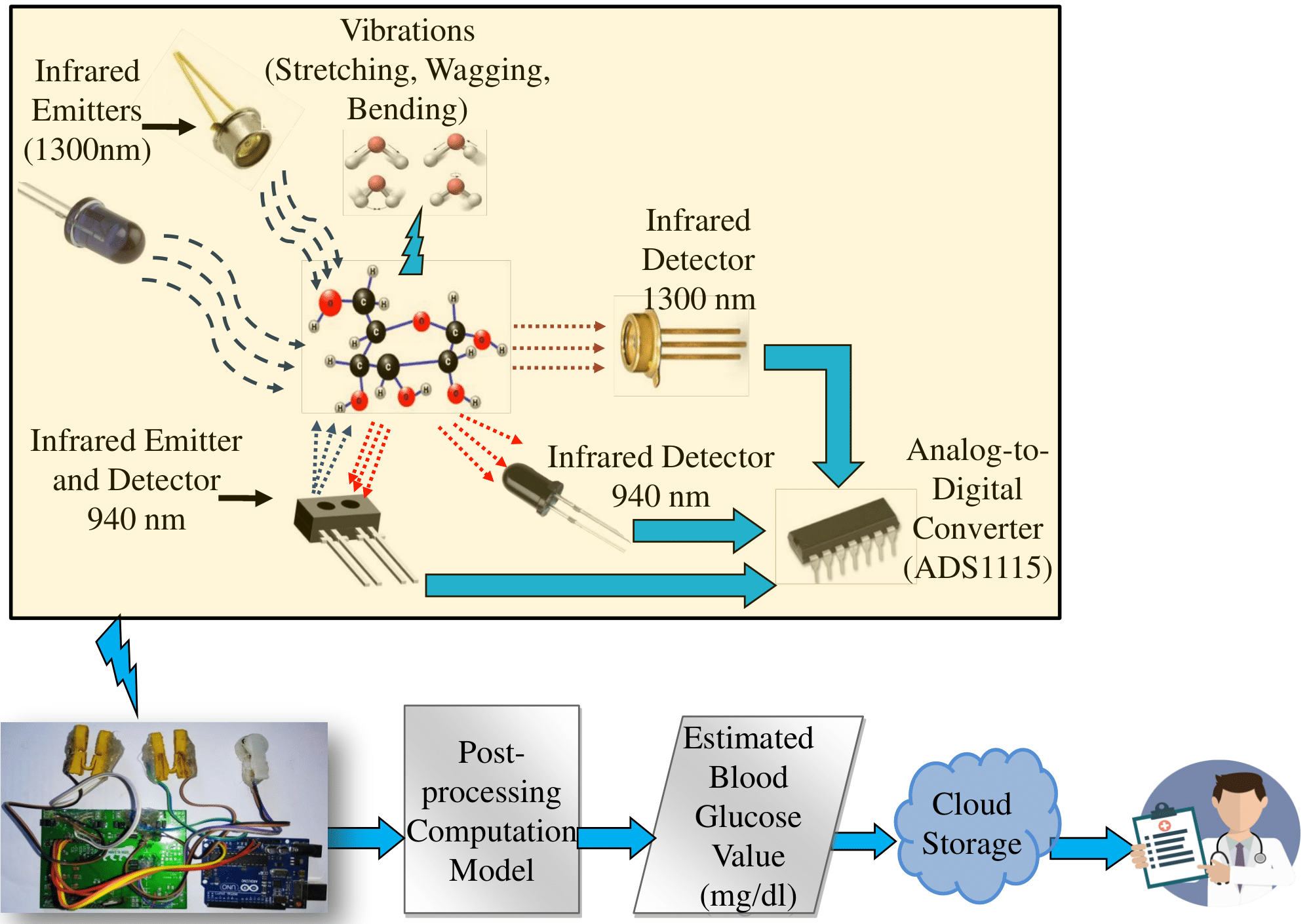}
	\caption{A conceptual overview of iGLU \cite{Jain_IEEE-MCE_2020-Jan_iGLU1}.}
	\label{cap}
\end{figure}

This article introduces an edge-device called ``Intelligent Glucose Meter'' (i.e. iGLU) for noninvasive, precise, painless, low-cost continuous glucose monitoring at the user-end and stores the data on the cloud in an IoMT framework. 
A non-invasive device has been proposed with a precise and low-cost solution. 
The proposed device is also integrated with IoMT where the data is accessible to the caretaker for point of care. The device will be portable after packaging to use everywhere. The device is fast operated and easy to use for smart healthcare. The flow of the proposed iGLU is represented in Figure \ref{cap}.

The contributions of this article to advance the state-of-art in smart healthcare include the following: 
\begin{enumerate}
	\item
	A novel accurate non-invasive glucometer (iGLU) by judiciously using short NIR waves with absorption and reflectance of light using specific wavelengths (940 and 1300 nm) has been introduced. The wavelengths are judiciously selected after experimental analysis which has been done in material research center MNIT, Jaipur (India). 
	\item
	A novel accurate machine learning-based method for glucose sensor calibration has been presented with calibrated and validated healthy, prediabetic and diabetic samples. 
	
	\item 
	The analysis method has been proposed based on
	multiple regression techniques for precise measurement
	of blood-glucose content.
	
	\item 
	A glucose monitoring device has been designed with proper biasing of sensors for calibration using an optimized regression model to validate and test the subjects.
	
	\item 
	With the active support from the diabetic center and hospital, real-life experimental validation has been done directly from human blood.
	
	\item
	The proposed non-invasive blood glucose measurement device has been integrated into IoMT framework for data (blood glucose values) storage, patient monitoring and treatment on proper time with cloud access by both the patient and doctor.
\end{enumerate}

\subsection{Why Short-Wave NIR Spectroscopy for Noninvasive Glucose Monitoring?}

\subsubsection{PPG versus NIR}
\begin{figure}[htbp]
	\centering
	\includegraphics[width=0.99\textwidth]{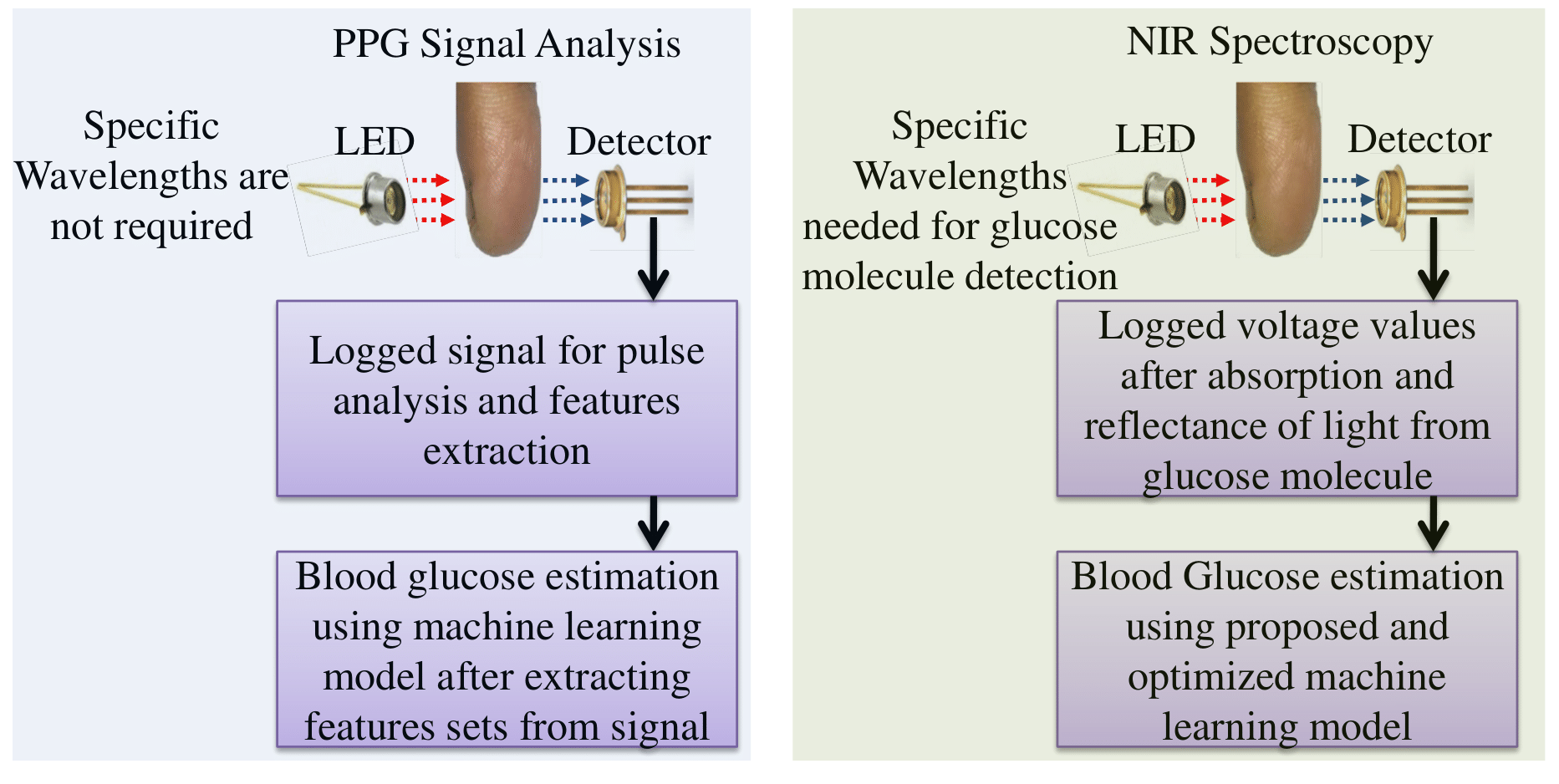}
	\caption{PPG signal versus NIRS based glucose estimation \cite{Jain_IEEE-MCE_2020-Jan_iGLU1}.}
	\label{NIRvsPPG}
\end{figure}

PPG signal can be used to detect the blood volume changes after light absorption by tissue \cite{habbu2019estimation}. Blood volume changes detected by the light detector which is caused by pressure pulse \cite{monte2011non}. The change in light intensity will be varied according to changes in blood volume. Hence, PPG signal analysis is not based on the principle of glucose molecule detection. Figure \ref{NIRvsPPG} shows the differences between PPG versus NIR. Therefore, specific wavelengths are not required for glucose estimation. Hence, iGLU is more precise compared to the PPG signal analysis based system for glucose measurement. An algorithm has been developed from the PPG signal for blood glucose monitoring \cite{paul2012design}. PPG signal is logged from a patient and different parameters which are correlated with the blood glucose value in the body. These parameters are extracted. Clinical parameters of that patient are also received. These are trained using various ML models and these trained models are analyzed for precise glucose measurement \cite{philip2017continous}. PPG signals from two classes (healthy and diabetic subjects) have been taken to predict the Auto-Regressive Moving Average (ARMA) models parameters. 70 healthy and the diabetic classes have been taken for analysis the parameters. Averaging of the predicted ARMA parameters have been done for each class, which leads to a different representative model for each class \cite{karimipour2009diabetic}. An FPGA based smart system is proposed for non-invasive Glucose measurement Using PPG signal with Online Correction of Motion Artifact \cite{ramasahayam2017fpga}. Non-invasive glucose monitoring is proposed with moving average filter using PPG signal in reflectance mode \cite{cruz2019application}. Non-invasive glucose estimation has also been done using smartphone PPG signals. They have used kNN classifier to optimized the computation model \cite{zhang2019non}. Experimental investigations have been done for blood glucose measurement using side scattered finger- PPG signals \cite{yamakoshi2017side}. Preliminary investigations have also been done for glucose measurement \cite{yamakoshi2015integrating}.

\subsubsection{Long-Wave versus Short-Wave NIR Spectroscopy}

\begin{figure}[htbp]
	\centering
	\includegraphics[width=0.70\textwidth]{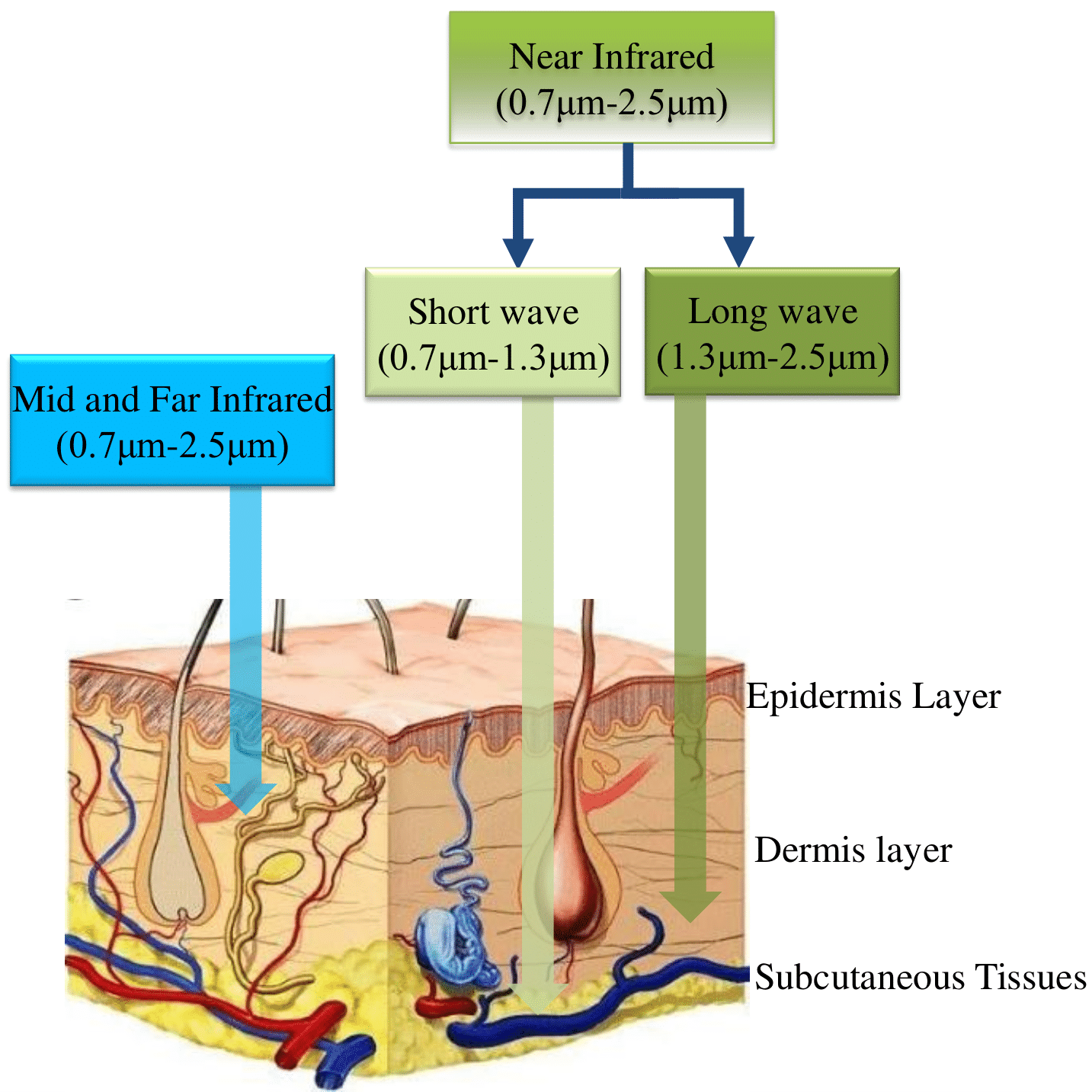}
	\caption{Representation of penetration depth in NIR and FIR region}
	\label{NIRvsFIR}
\end{figure}

Optical approach based detection has been found with fewer limitations in terms of precise measurement. In this way, FIR (Far infra-red) wave for optical detection has been considered for glucose measurement. In FIR region, the resonance between CH and OH are known as first overtone resonance. Glucose molecule detection is more precise using NIR long wave for $in vitro$ testing. In this way, $in vitro$ glucose measurement has been done using tunable mid-infrared laser spectroscopy which is combined with fiber-optic sensor \cite{yu2014vitro}. Error analysis of mid-infrared sensors has been done using multivariate calibration model for continuous glucose monitoring \cite{goodarzi2016selection}. Calibration, validation and testing have been done for $in vitro$ measurement \cite{goodarzi2015multivariate}. In this way, more informative wave bands have been identified in mid-infrared region for glucose molecule detection. But, the main disadvantage is shallow penetration in FIR region (first overtone) compared short wave NIR region. The glucose molecule detection is more precise in the FIR region \cite{sharma2013efficient}. Therefore, small NIR wave is preferred for glucose detection which is represented in Fig. \ref{NIRvsFIR}.
The precision level in non-invasive blood glucose measurement has been explored by NIR spectroscopy with specific wavelengths \cite{uwadaira2010factors}. CH vibration has been found near 920 nm without conflicting of water molecule vibration. Infrared spectra of sucrose, fructose and glucose have been presented for non-invasive measurement \cite{uwadaira2010factors}. In this way, 940 nm wavelength has been chosen for glucose molecule detection \cite{haxha2016optical}. Stretching of glucose molecule has also been examined in NIR region \cite{golic2003short}. Validation of glucose absorption has also been donein range of 1300-1350 nm \cite{zhang2013discussion}. Then, 1300 nm wavelength has also been chosen for non-invasive glucose measurement \cite{Song2015}. After literature survey and real-time experimental analysis, the specific wavelengths have been chosen for glucose measurement.

\section{Proposed Non invasive Blood Glucose Measurement Device (iGLU)}
\label{Sec:Proposed-Work}

\begin{figure}[htbp]
	\centering
	\includegraphics[width=0.80\textwidth]{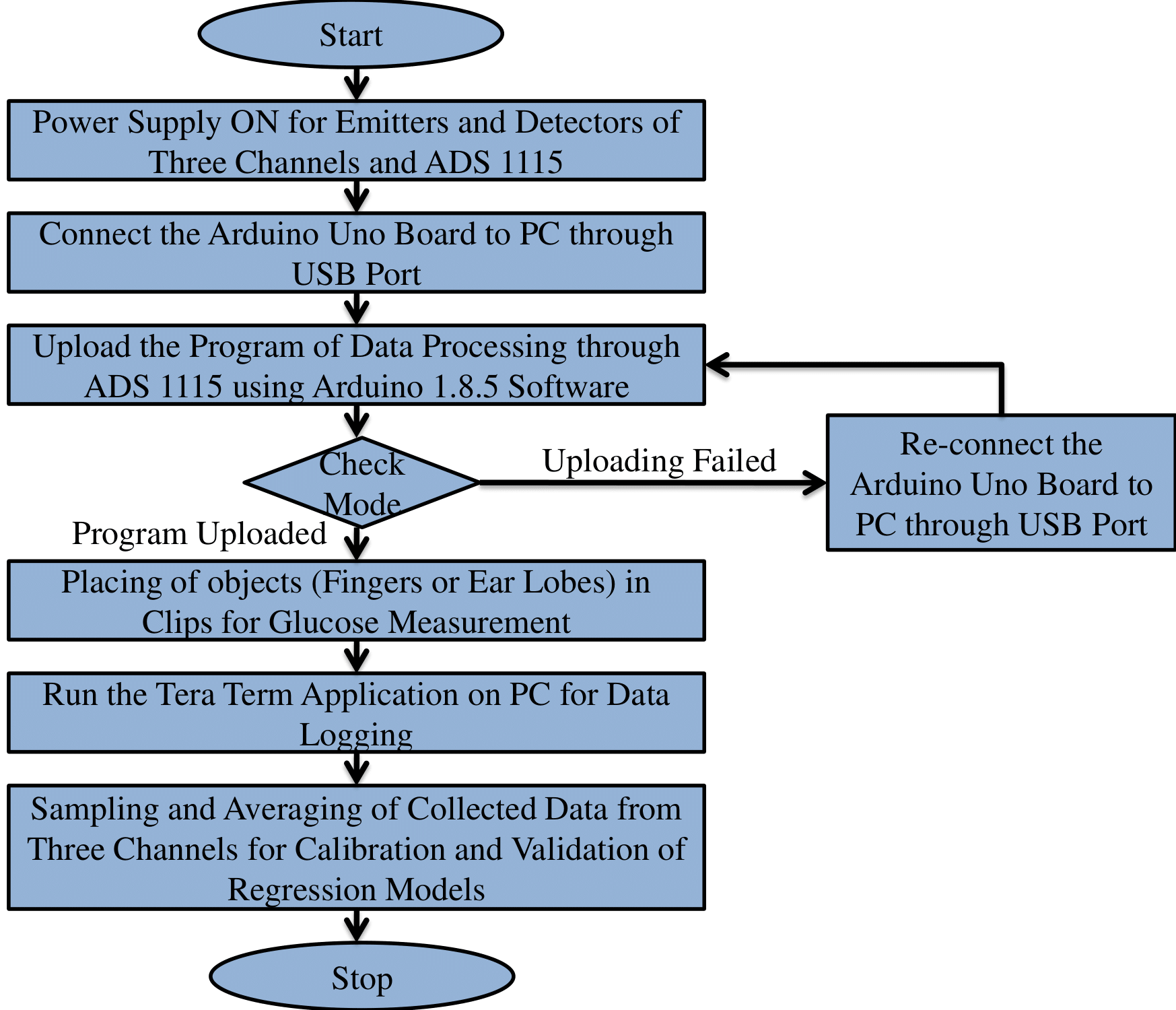}
	\caption{Process flow data acquisition for proposed device (iGLU) \cite{Jain_IEEE-MCE_2020-Jan_iGLU1}.}
	\label{pseudocode}
\end{figure}

The proposed device based on NIR spectroscopy with multiple short wavelengths is designed and implemented using three channels. Each channel is embedded with emitter and detector of specific wavelength for optical detection. The data is collected and serially processed by 16 bit ADC with sampling rate of 128 samples per second. The logged data is calibrated and validated thorough existing regression techniques to analyse the optimized model \cite{jain2019precise}. The pseudocode of data acquisition for proposed iGLU is represented in process flow which is shown in Fig. \ref{pseudocode}.

Independent samples have also been taken for testing to validate the iGLU. The blood glucose concentration values have been stored on cloud using open IoT platform. The data can be accessed by patients and doctor. On basis of blood glucose values, the treatment can be provided at remote location.

\subsection{The Approach for Glucose Molecule Detection}

Glucose molecule vibrates according to its atomic structure at specific wavelengths. It is analyzed that absorbance and reflectance are sharper and stronger in short wave NIR region \cite{Uwadaira2016}. Zhang et.al. analyzed the absorption peak of glucose spectra at 1314 nm \cite{zhang2013discussion}. Song et.al. implemented the non-invasive blood glucose measurement using 850, 950 and 1300 nm \cite{Song2015}. Haxha et.al. proposed 940 nm wavelength for detection of glucose molecule \cite{haxha2016optical}. Golic et.al. elaborated near infrared spectra of sucrose, glucose and fructose with $CH_{2}$, CH and OH stretching at 930 nm, 960 nm and 984 nm respectively \cite{golic2003short}. Beckers also examined glucose apsorptions in short wave NIR region (800-1400 nm) \cite{Beckers}. Most informative glucose absorption peaks reported are in wave bands 920-960 nm and 1300-1320 nm according to the literature \cite{Song2015}.

\subsection{Proposed Module for Data Acquisition}

\begin{figure}[htbp]
	\centering
	\includegraphics[width=0.99\textwidth]{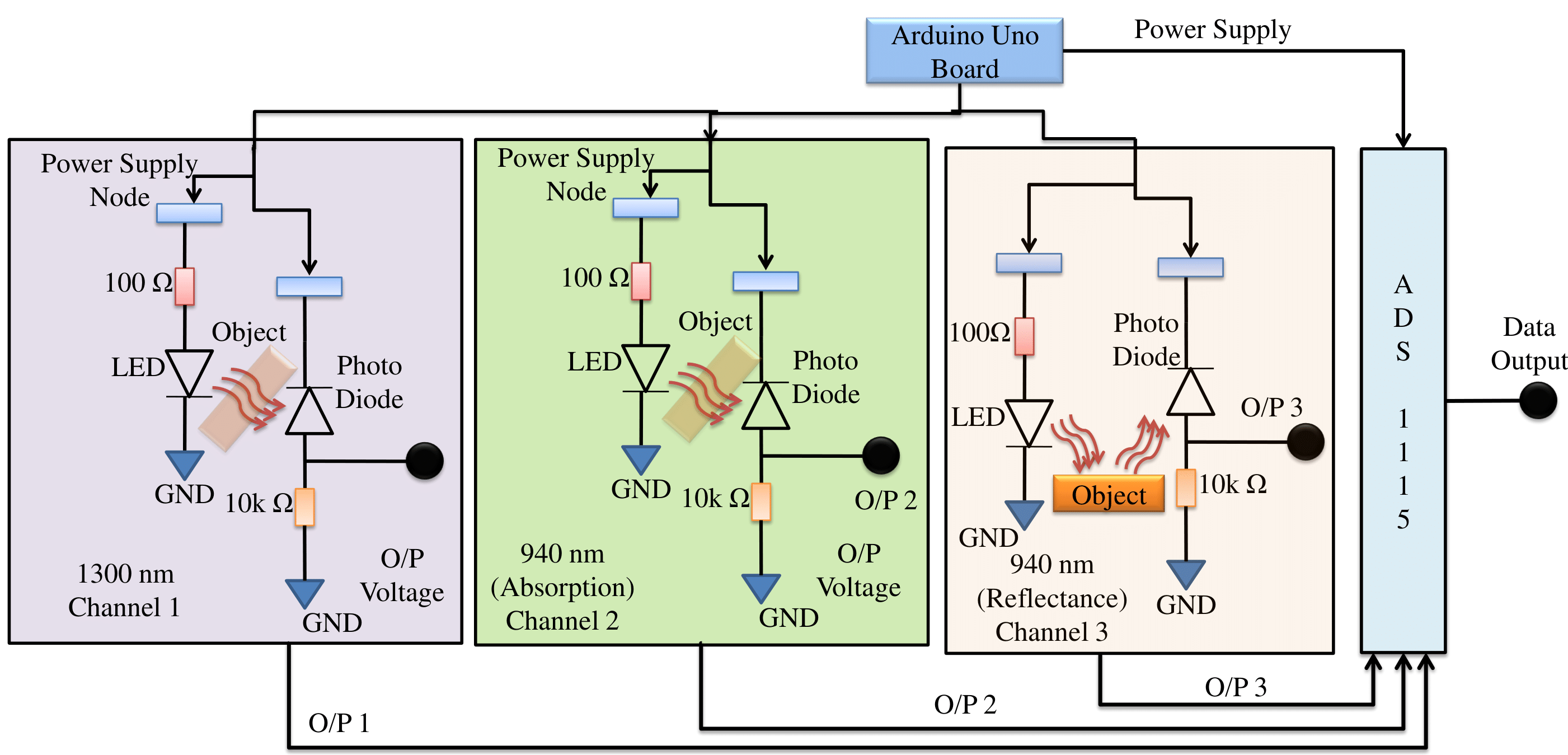}
	\caption{Circuit topology of proposed device (iGLU).}
	\label{circuit}
\end{figure}

Proposed iGLU uses NIR spectroscopy to improve the accuracy. A 2-Layer PCB has been developed to embed infra-red emitters (MTE1300W -for 1300 nm, TSAL6200 -for 940 nm, TCRT1000 -for 940 nm) and detectors (MTPD1364D -for 1300 nm, 3004MID -for 940 nm, TCRT1000 -for 940 nm). The hardware is designed for data acquisition from emitters, detectors and ADC with 5V DC supply \cite{Jain2016, Jain2017}. According to the emitters and detectors, compatible passive components have been chosen. Architecture of glucose sensing is shown in Figure \ref{circuit} \cite{Jain2016,Jain2017}. 
Detectors with daylight blocking filters are packaged and not affected by sweat \cite{jain2014analyzing}. ADS 1115 with 860 SPS, 16 bit, $I^2C$ compatible and single ended is controlled through microcontroller ATmega328P and used to convert the data (in Volts) from all channel in decimal form \cite{jain2018full,jain2014analysis}. The noise power and signal-to-noise ratio (SNR) have also been found 0.08 and 25.2 dB, respectively, which show the minimum noise level.

The voltage and current ratings of emitters and detectors are presented in Table \ref{rating}.

\subsection{A Specific Prototype of the iGLU}

Absorption and reflectance at 940 nm and absorption at 1300 nm are implemented for detection of the glucose molecules. The detector's voltage depends on received light intensity. After placing the fingertip between emitter and detector, the voltage values are logged. Change in light intensity depends upon glucose molecule concentration. During experiments, blood glucose is measured through the invasive device standard diagnostics (SD) check glucometer for validation of the non-invasive results. The reading is taken as referenced blood glucose values (mg/dl). During the process, optical responses through detectors have been collected from 3 channels simultaneously. During measurement, the channels data is collected in the form of voltages from 3 detectors. These collected voltages correspond to referenced blood glucose concentration. These voltage values are converted into decimal form using 4-channel ADS 1115 (Texas Instruments) ADC \cite{ jain2014analysis}. Coherent averaging has been done after collection of responses. Specification of a iGLU prototype are presented in Table \ref{rating}.

\begin{table}[htbp]
	\caption{Specification of iGLU prototype}
	\label{rating}
	\centering
	\begin{tabular}{p{3.5cm}p{1.9cm}p{2.3cm}p{2.3cm}}
		\hline
		&Channel 1&Channel 2&Channel 3\\
		&	\multicolumn{3}{c}{Measured (Ideal)}\\
		\hline
		\hline
		Arduino Supply &4.95V&4.96V& 4.95V\\
		&(5V)&(5V)&(5V)\\
		\hline
		Forward Voltage&0.96V&1.42V& 1.40V\\
		(Emitter)&(1.1V)&(1.5V)&(1.5V)\\
		\hline
		Forward Current&53.4mA&52.8mA&52.9mA\\
		(Emitter)&(100mA)&(60mA)&(60mA)\\
		\hline
		Reverse Voltage&4.25V&4.16V&4.25V\\
		(Detector)&(5V)&(5V)&(5V)\\
		\hline
		Output Current&0.45mA&0.5mA&0.52mA\\
		(Detector)&(1mA)&(1mA)&(1mA)\\
		\hline
		Measurement range&3.2-4.68V&0.8-4.7V&0.5-4.7V\\
		\hline
		Specific Wavelength&1300nm&940nm&940nm\\
		\hline
		Spectroscopy&Absorption&Absorption&Reflectance\\
		\hline
	\end{tabular}
\end{table}

\begin{figure}[htbp]
	\centering
	\includegraphics[width=0.4\textwidth,angle=-90]{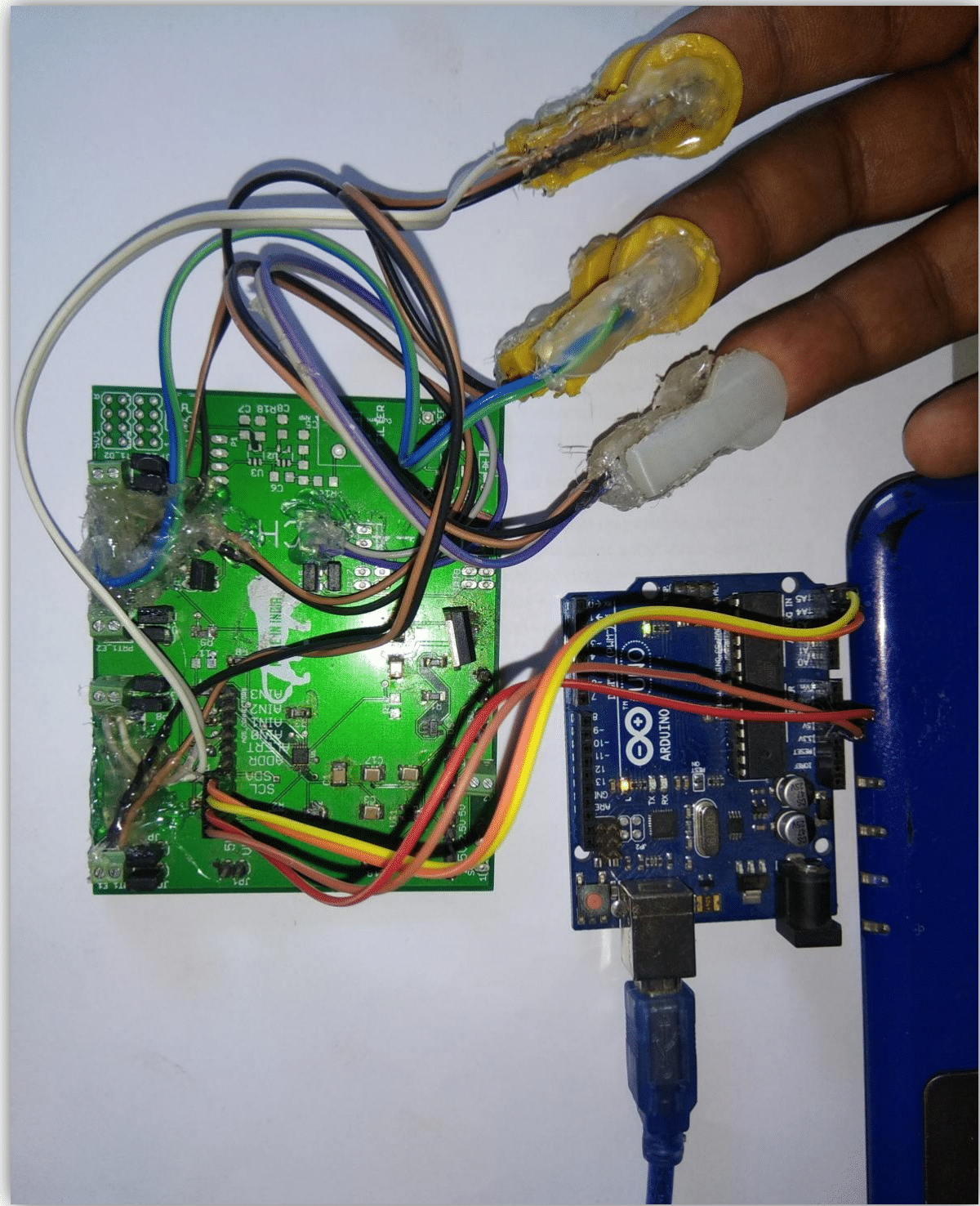}
	\caption{Prototype view of proposed device (iGLU).}
	\label{fig:Prototyping_Glucose_Measurement_System}
\end{figure}

The prototype view of proposed iGLU is shown in Figure \ref{fig:Prototyping_Glucose_Measurement_System}. The data is collected after fixing three fingers in the free space of pads. The pads are designed in such a way that emitters and detectors are placed beneath the surfaces of pads. Because of this, there will be enough free spaces between the object and sensors (emitters and detectors). Hence, the probability of a faulty measurement is minimized. 

\section{Proposed Machine-Learning (ML) based Methods for iGLU Calibration}
\label{Sec:Calibration}

\begin{figure}[htbp]
	\centering
	\includegraphics[width=0.75\textwidth]{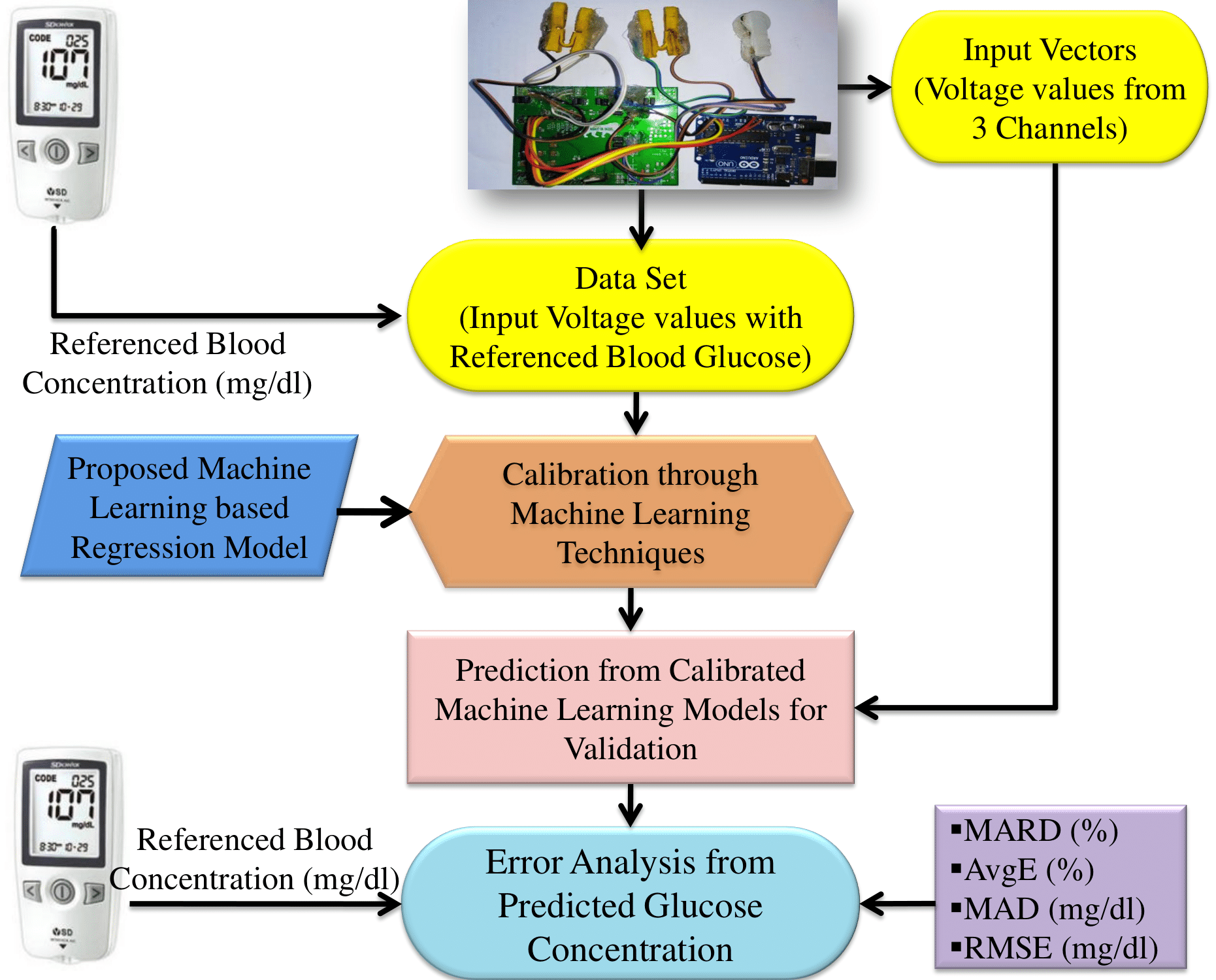}
	\caption{The process flow of calibration and validation of proposed device (iGLU) \cite{Jain_IEEE-MCE_2020-Jan_iGLU1}.}
	\label{ML_flow}
\end{figure}

Regression models (RM) are calibrated to analyze the optimized computation model for blood glucose estimation. The detector's outputs from three channels are logged as input vectors for prediction of glucose concentrations. The calibrated models are used to predict the blood glucose concentrations for validation. The collected data from the samples are required to be converted in the form of estimated blood glucose concentration values. It is necessary to design an optimized kernel for precise measurement of the predicted glucose concentration value. 97 samples are taken for device calibration which includes prediabetic, diabetic and healthy samples. The baseline characteristics of samples for calibration is represented in Table \ref{dataset}. The proposed process flow of calibration and validation is shown in Fig. \ref{ML_flow}.

\subsection{Regression Models for Prediction of Blood Glucose from Detector's Output Voltages}

Various machine learning-based regression models have been analyzed to get an optimized regression method. In this way CLS (classical least squares), PLS (partial least squares), NPLS (multiway PLS), MLR (multiple linear regression), SVR (support vector regression), logistic regression, PCR (principal component regression), MPR (multiple polynomial regression) and neural network fitting model have been tried to find an optimized model for precise blood glucose measurement.
To analyze the accuracy of the predicted glucose concentration,  mean absolute deviation ($MAD$), mean absolute relative difference ($mARD$), and root mean square error $RMSE$ are calculated using Eqn. (\ref{mad}), (\ref{mard}), (\ref{rmse}):
\begin{eqnarray}
\label{mad}
MAD & = & \frac{1}{n}\sum\limits_{i=1}^{n}\left |{BG_{Est}-BG_{Ref}} \right |,\\
\label{mard}
mARD & = & \frac{1}{n}\sum\limits_{i=1}^{n}\left | \frac{BG_{Est}-BG_{Ref}}{BG_{Ref}} \right |\times 100\\
\label{rmse}
RMSE & = & \sqrt{\frac{1}{n}\sum_{i=1}^{n}\left | (BG_{Ref}-BG_{Est}) \right |^2}\\
\label{avg}
AvgE & = & \frac{1}{n}\sum\limits_{i=1}^{n}\left | \frac{BG_{Est}-BG_{Ref}}{BG_{Est}} \right |\times 100
\end{eqnarray}
In the above expression, $BG_{Est}$ and $BG_{Ref}$ are estimated and reference blood glucose concentration respectively.
From Eqn. (\ref{avg}), $AvgE$ is an average (mean) error or deviated error from blood glucose concentration values, $n$ is total samples.

\begin{table}[htbp]
	\caption{Baseline characteristics of collected samples for calibration.}
	\label{dataset}
	\centering
	\begin{tabular}{ll}
		\hline
		Samples Basic & Brief Explanation\\
		Characteristics & of Samples for Calibration\\
		\hline
				\hline
		Age (Years) & Prediabetic\\
		Male:-   22-65 & Male:-   18\\
		Female:- 26-75 & Female:- 13\\
		\hline
		Age (Years) &  Diabetic\\
		Male:-   30-68 & Male:-   16\\
		Female:- 30-73 & Female:- 14\\
		\hline
		Age (Years) & Healthy\\
		Male:-   22-65 & Male:-   19\\
		Female:- 17-70 & Female:- 17\\
		\hline
		Age (Years) & Total Samples\\
		Male:-   22-77 & Male:-   53\\
		Female:- 17-75 & Female:- 44\\
		\hline
	\end{tabular}
\end{table}

\subsection{Multiple Polynomial Regression (MPR) Model for Estimation of Blood Glucose}

Multiple polynomial regression (MPR) is applied to design the model for blood glucose computation. Using multiple polynomial regression, the model explores the relationship between multiple independent variables and a dependent variable by fitting a computation model to observe the estimated value. Each value of explanatory variable x is associated with a value of response variable y. In the proposed system, multiple polynomial regression line for three channels variable $x_{1}$, $x_{2}$ and $x_{3}$ is represented as independent variables and estimated glucose (mg/dl) is the dependent variable $y$.
According to the observation of these 97 samples proposed multiple polynomial regression model (MPR) can be represented as a polynomial of degree 3:
\begin{eqnarray}
y=a_{1}x_{1}^{3}+a_{2}x_{2}^{3}+a_{3}x_{3}^{3}+a_{4}x_{1}^{2}x_{2}+a_{5}x_{1}^{2}x_{3}+a_{6}x_{1}x_{2}^{2}+a_{7}x_{1}x_{3}^{2}+a_{8}x_{2}^{2}x_{3}+a_{9}x_{2}x_{3}^{2}+a_{10}x_{1}^{2}+a_{11}x_{2}^{2}\nonumber\\+a_{12}x_{3}^{2}+a_{13}x_{1}x_{2}x_{3}+a_{14}x_{1}x_{2}+a_{15}x_{1}x_{3}+a_{16}x_{2}x_{3}+a_{17}x_{1}+a_{18}x_{2}+a_{19}x_{3}+\epsilon
\label{1}
\end{eqnarray}
where, $a$, $b$, and $c$ are regression coefficients and $\epsilon$ is residuals which can be called as a random error in prediction. Similarly, MPR with polynomial degree 4 has been used to analyze the optimized model for glucose measurement. 
During analysis, the data from two channels (absorption and reflectance of light at 940 nm wavelength) has been collected, then this MPR based calibration (RM1) with polynomial degree 3 as well as degree 4 are used to relate the dependent variables to the independent variables. The dependent variable is referenced blood glucose concentration (mg/dl) and independent variables are the voltage values of detectors.
Now, the data from another two channels (absorption of light at 940 nm and 1300 nm) is collected and modelled using multiple polynomial kernel based calibration (RM2) with polynomial degree 3 and 4.
Now, the data from another two channels (reflectance of light at 940 nm and absorption of light at 1300 nm) is collected and modelled using MPR kernel-based calibration (RM3) with polynomial degree 3 and 4.
Now, the data from all three channels is collected and modelled using multiple polynomial kernel based calibration (RM4) with polynomial degree 3. Pearson's correlation coefficient (R) is found 0.98.
mean absolute deviation ($MAD$), mean absolute relative difference ($mARD$), root mean square error $RMSE$ and average error ($AvgE$) are calculated which have been summarized in Table \ref{comb1} and Table \ref{comb}.

\begin{table}[!h]
	\caption{Statistical Analysis of possible combinations of spectroscopy techniques using MPR kernel based calibration with polynomial degree 3.}
	\label{comb1}
	\centering
	\begin{tabular}{cccccc}
		\hline
		\textbf{}&\textbf{$R^{2}$}&\textbf{$mARD$}&\textbf{$AvgE$}&\textbf{$MAD$}&$RMSE$\\
		\textbf{}&value&(\%)&(\%)&(mg/dl)&(mg/dl)\\
		\hline		\hline
		RM1&0.42&30.74&24.78&53.22&73.68\\
		\hline
		RM2&0.46&28.58&22.33&47.61&66.94\\
		\hline
		RM3&0.43&29.76&23.70&50.83&69.73\\
		\hline
		\textbf{RM4}&\textbf{0.81}&\textbf{4.66}&\textbf{4.61}&\textbf{7.55}&\textbf{11.95}\\
		\hline
	\end{tabular}
\end{table} 

\begin{table}[!h]
	\caption{Statistical Analysis of possible combinations of spectroscopy techniques using MPR kernel based calibration with polynomial degree 4.}
	\label{comb}
	\centering
	\begin{tabular}{cccccc}
		\hline
		\textbf{}&\textbf{$R^{2}$}&\textbf{$mARD$}&\textbf{$AvgE$}&\textbf{$MAD$}&$RMSE$\\
		\textbf{}&value&(\%)&\textbf{(\%)}&(mg/dl)&(mg/dl)\\
		\hline		\hline
		RM1&0.32&43.68&32.81&59.66&75.22\\
		\hline
		RM2&0.28&30.21&27.68&52.77&72.11\\
		\hline
		RM3&0.33&32.34&34.32&57.78&70.21\\
		\hline
		\textbf{RM4}&\textbf{0.56}&\textbf{10.61}&\textbf{10.69}&\textbf{10.23}&\textbf{17.46}\\
		\hline
	\end{tabular}
\end{table} 
 
From Table \ref{comb1} and \ref{comb}, it is observed that the MPR kernel-based calibration with polynomial degree 3 has better results compared to same kernel-based calibration with polynomial degree 4 for a possible combination of three channels. Hence, RM4 (MPR3) is considered as an optimized kernel for blood glucose estimation.

\subsection{Deep Neural Network (DNN) Modeling for Blood Glucose Prediction}

\begin{figure}[htbp]
	\centering
	\includegraphics[width=0.85\textwidth]{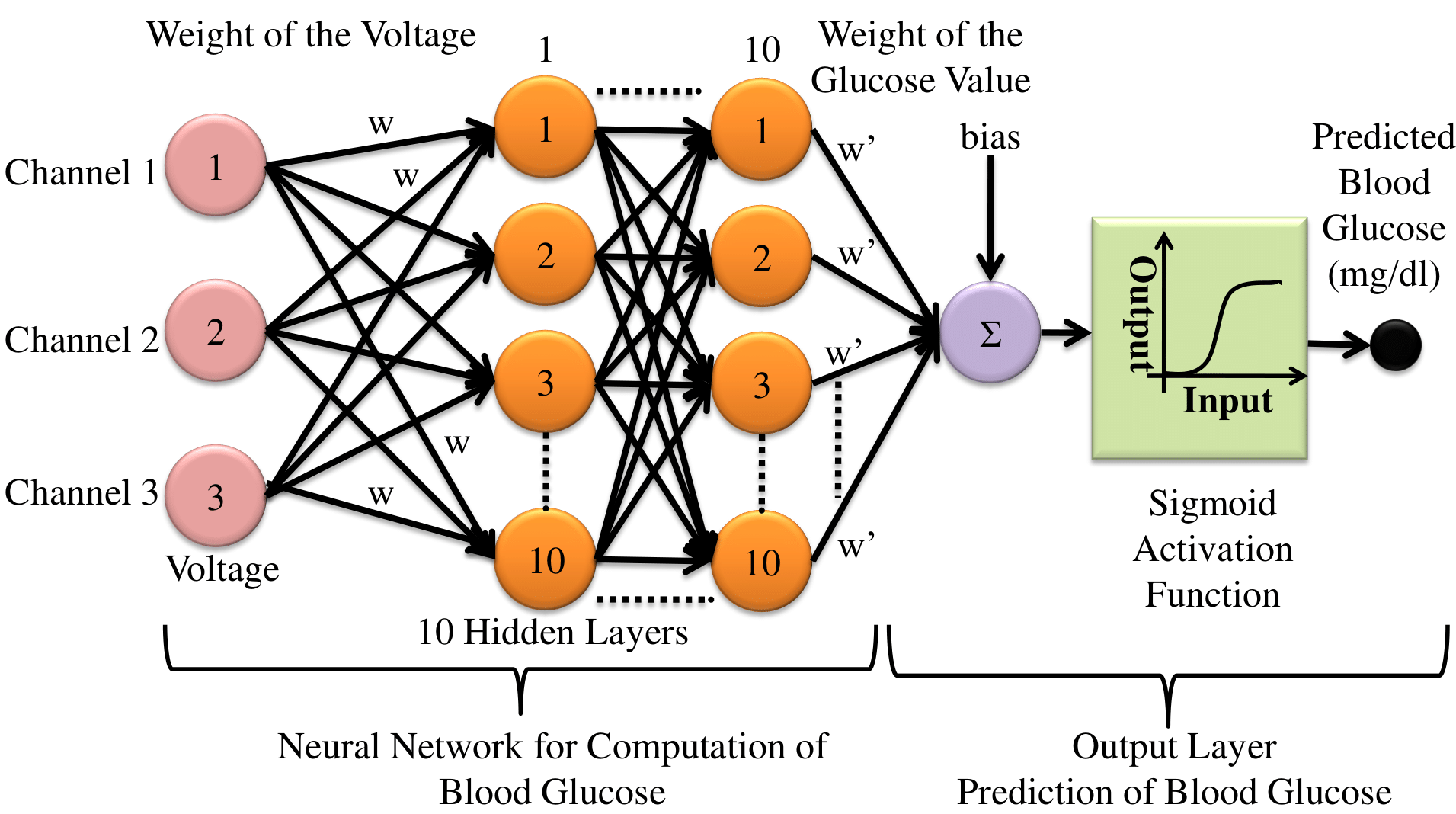}
	\caption{The Deep Neural Network (DNN) model integrated in the iGLU \cite{Jain_IEEE-MCE_2020-Jan_iGLU1}.}
	\label{dnnfit}
\end{figure}

As a 2nd method, the deep neural network (DNN) model has been explored for the glucose level predication in the current paper \cite{Sundaravadivel_IEEE-TCE_2018-Aug}. Proposed DNN uses sigmoid activation functions and has been trained through Levenberg-Marquardt backpropagation algorithm \cite{Song2015}. In the proposed model, 10 hidden neurons and 10 hidden layers are analyzed to estimate the precise blood glucose values. This model has been used to analyze the non-linear statistical data which is utilized to calibrate and validate the model for precise measurement. Here, the voltage values from three channels are used as inputs of the proposed DNN model. The predicted blood glucose values are formed through the modelling of three channels of voltage values. Weights of the voltage values correlate predicted glucose values to the channels data. The overall accuracy is improved using 10 hidden layers. The block diagram of the neural network fitting model is given in Fig. \ref{dnnfit}. 

\begin{figure*}[htbp]
	\centering
	\subfigure[Logistic]
	{\includegraphics[width=0.45\textwidth]{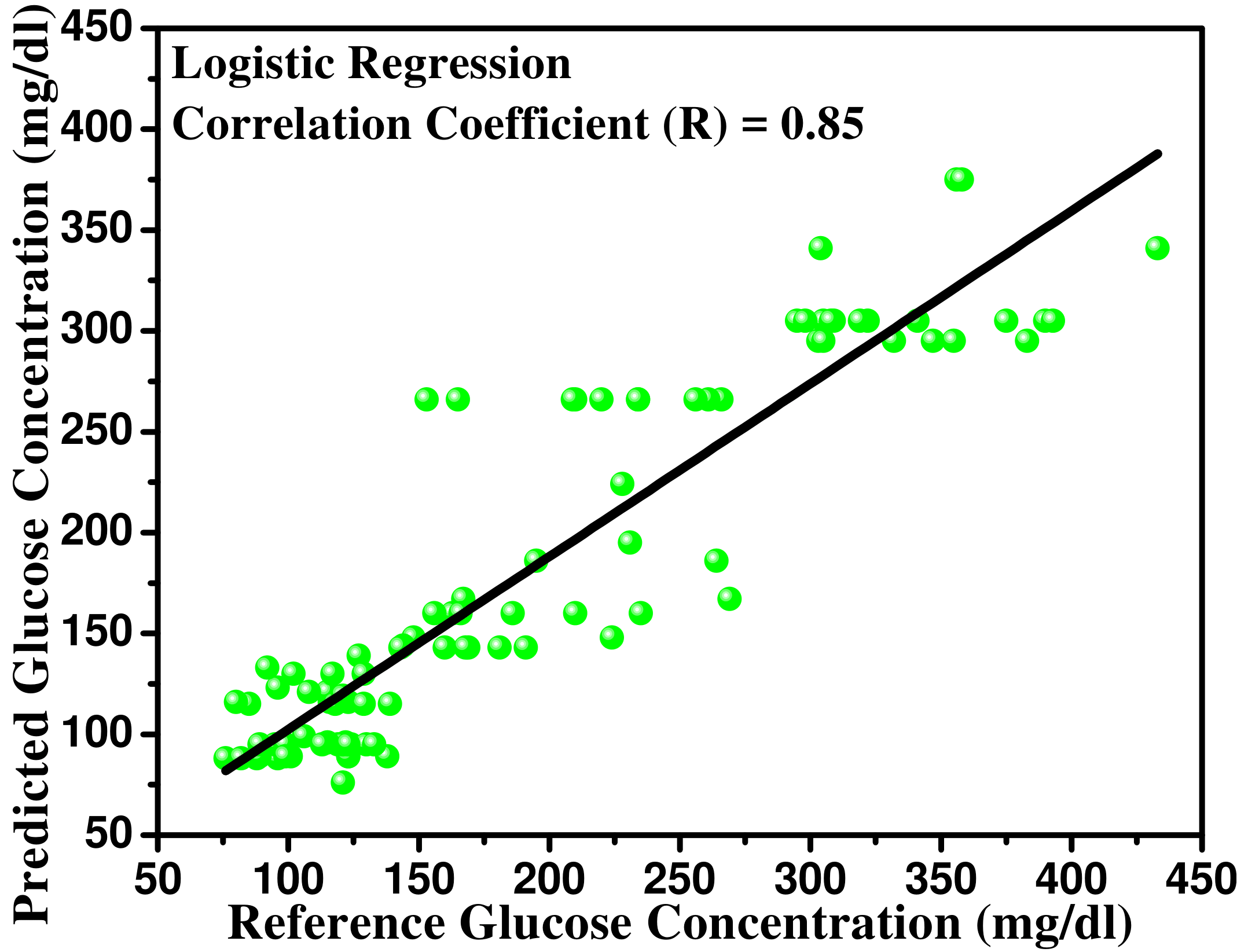}}\label{}
	\subfigure[SVR]
	{\includegraphics[width=0.45\textwidth]{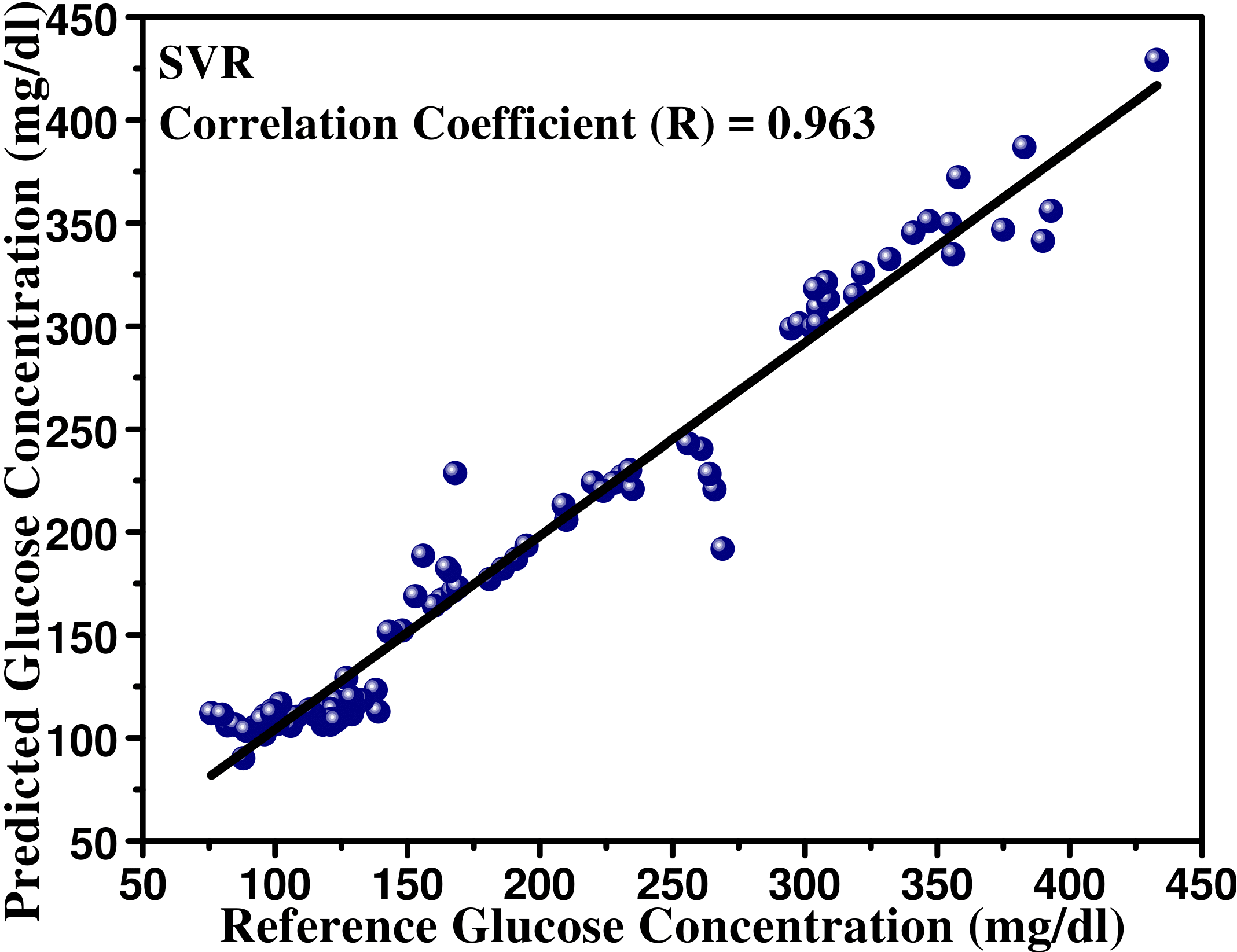}}\label{}
	\subfigure[DNN]
	{\includegraphics[width=0.45\textwidth]{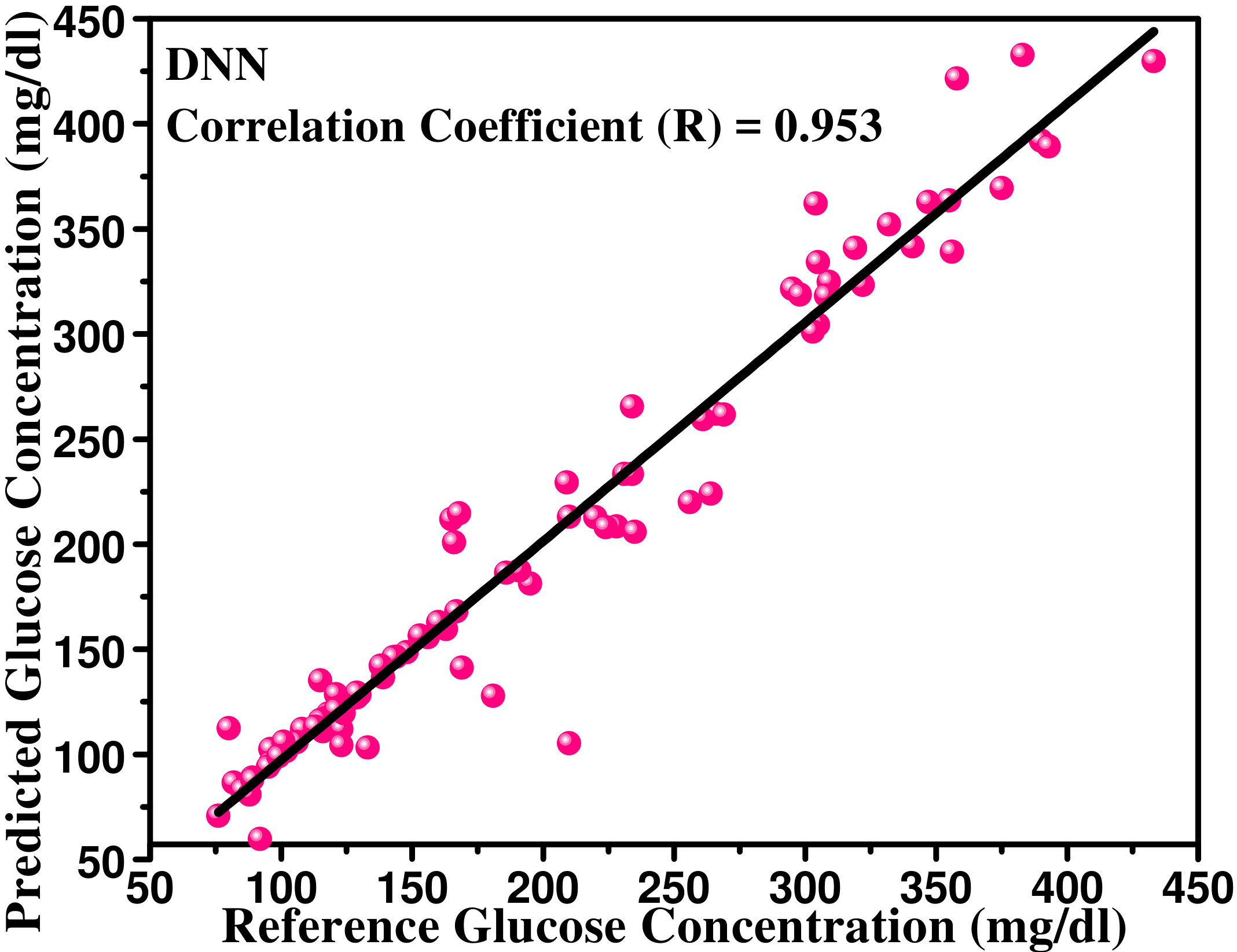}}\label{}
	\subfigure[MPR3 (RM4)]
	{\includegraphics[width=0.45\textwidth]{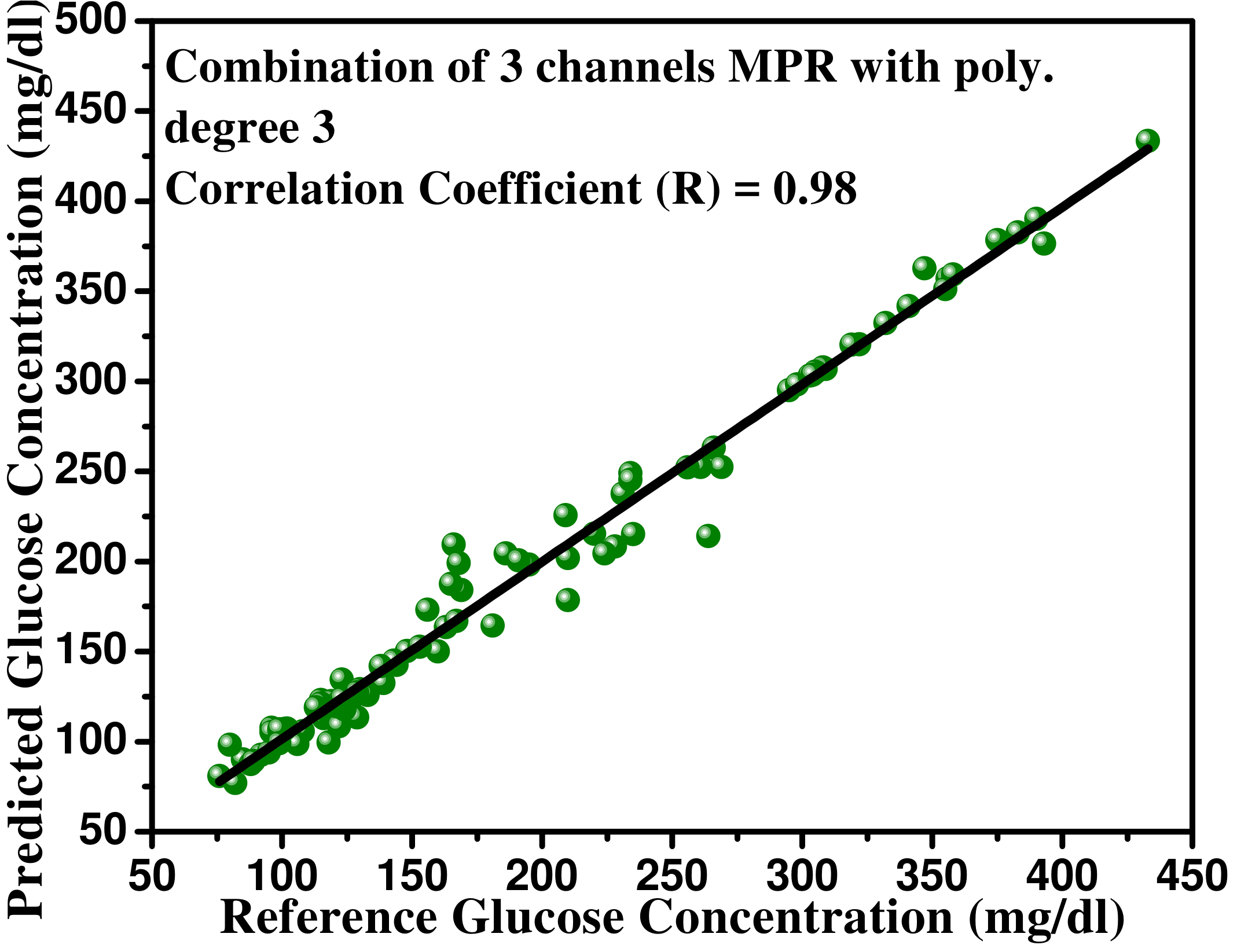}}\label{2}
	\caption{Correlation plots of predicted and referenced blood glucose concentration of proposed and existing regression models.} 
	\label{wave}
\end{figure*}

Pearson's correlation coefficient (R) is 0.953. The error analysis of calibrated machine learning models is represented in Table \ref{TBL:calibration}. The predicted and referenced blood glucose concentrations are represented in Fig. \ref{wave}(a)- Fig. \ref{wave}(d).

\begin{table}[htbp]
	\caption{Statistical Analysis of calibration of proposed model and existing techniques.}
	\label{TBL:calibration}
	\centering
	\begin{tabular}{ccccc}
		\hline
		Regression&$mARD$&$AvgE$&$MAD$&$RMSE$\\
		Model&\%&\%&mg/dl&mg/dl\\
		\hline
		\hline
		Logistic&15.11&16.63&27.78&39.06\\
		\hline
		SVR&8.03&7.71&12.34&18.04\\
		\hline
		DNN&6.65&7.30&12.67&21.95\\
		\hline
		\textbf{MPR3(RM4)}&\textbf{4.66}&\textbf{4.61}&\textbf{7.55}&\textbf{11.95}\\
		\hline
	\end{tabular}
\end{table}

\section{Validation of the Proposed Device}
\label{Sec:Experimental-Results}

\begin{figure}[!h]
	\centering
	\includegraphics[width=0.6\textwidth]{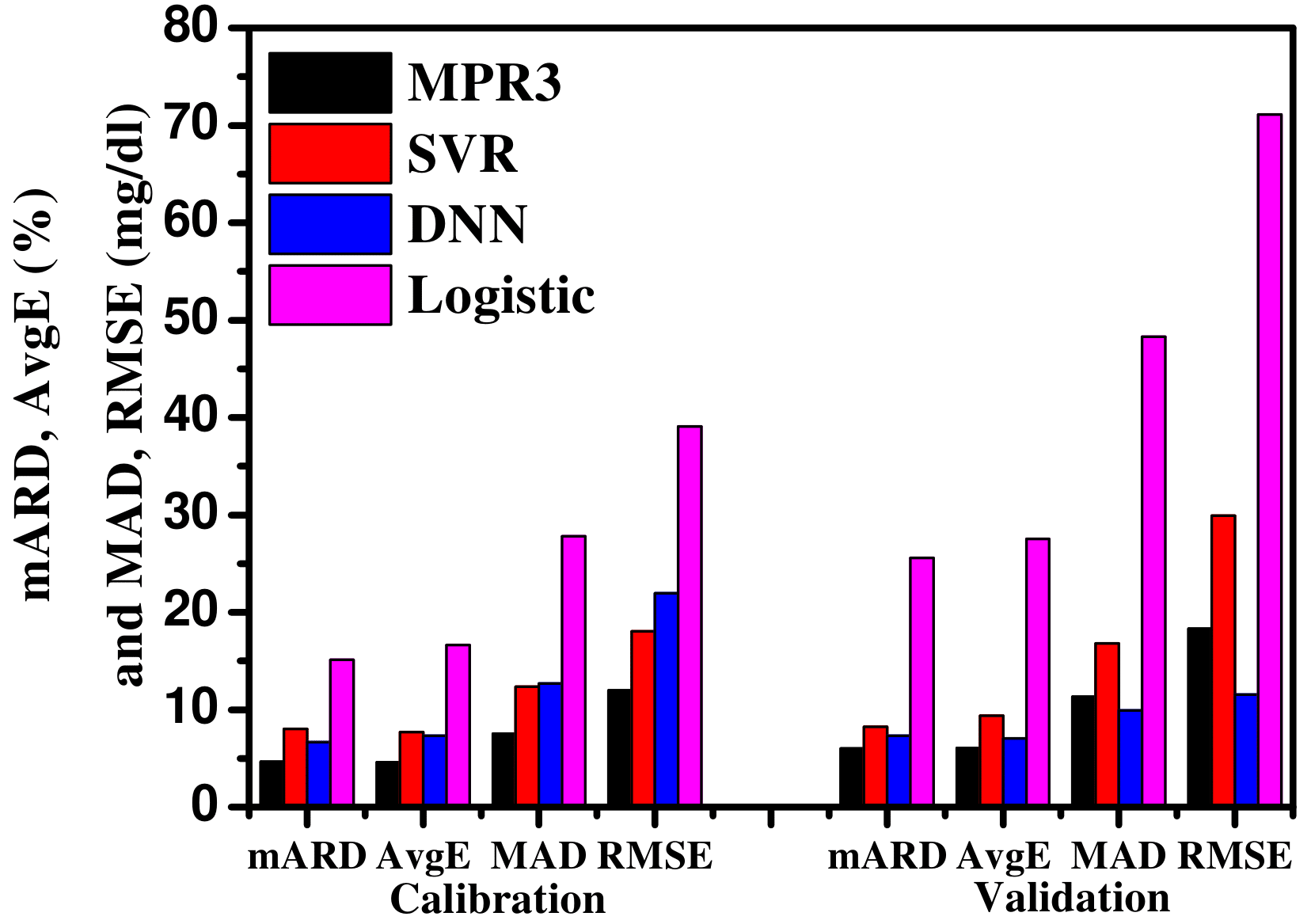}
\caption{Error analysis of validation of data using existing and proposed calibrated regression models.}
	\label{validation}
\end{figure}

After calibration of iGLU, 93 healthy, prediabetic and diabetic samples aged 17 to 75 are taken to validate and test the device iGLU following medical protocols. 64 males and 29 females samples are found during Collection of these 93 samples. All samples are taken in fasting, post-prandial and random modes. The baseline characteristics and error analysis is represented in Table \ref{dataset1} and \ref{val} respectively. 10-fold cross-validation has been performed to validate iGLU. According to graphical representation in Fig \ref{validation}, it is concluded that the measurement is more precise with the proposed combination of channels using MPR with polynomial degree 3. The resultant values also confirm the precise measurement of blood glucose level.
The graphical representation of error analysis for calibration and validation of iGLU is shown in Fig. \ref{validation}.

\begin{table}[htbp]
	\caption{Baseline characteristics of collected samples for validation and testing.}
	\label{dataset1}
	\centering
	\begin{tabular}{ll}
		\hline
		Samples Basic & Brief Explanation\\
		Characteristics & of Samples for Validation\\
		\hline
			\hline
		Age (Years) & Prediabetic\\
		Male:-   22-65 & Male:-   11\\
		Female:- 26-75 & Female:- 10\\
		\hline
		Age (Years) &  Diabetic\\
		Male:-   30-68 & Male:-   17\\
		Female:- 30-73 & Female:- 11\\
		\hline
		Age (Years) & Healthy\\
		Male:-   22-65 & Male:-   36\\
		Female:- 17-70 & Female:- 08\\
		\hline
		Age (Years) & Total Samples\\
		Male:-   22-77 & Male:-   64\\
		Female:- 17-75 & Female:- 29\\
		\hline
	\end{tabular}
\end{table}

\begin{figure*}[htbp]
	\centering
	\subfigure[Time interval of 5 minutes]{\includegraphics[width=0.42\textwidth]{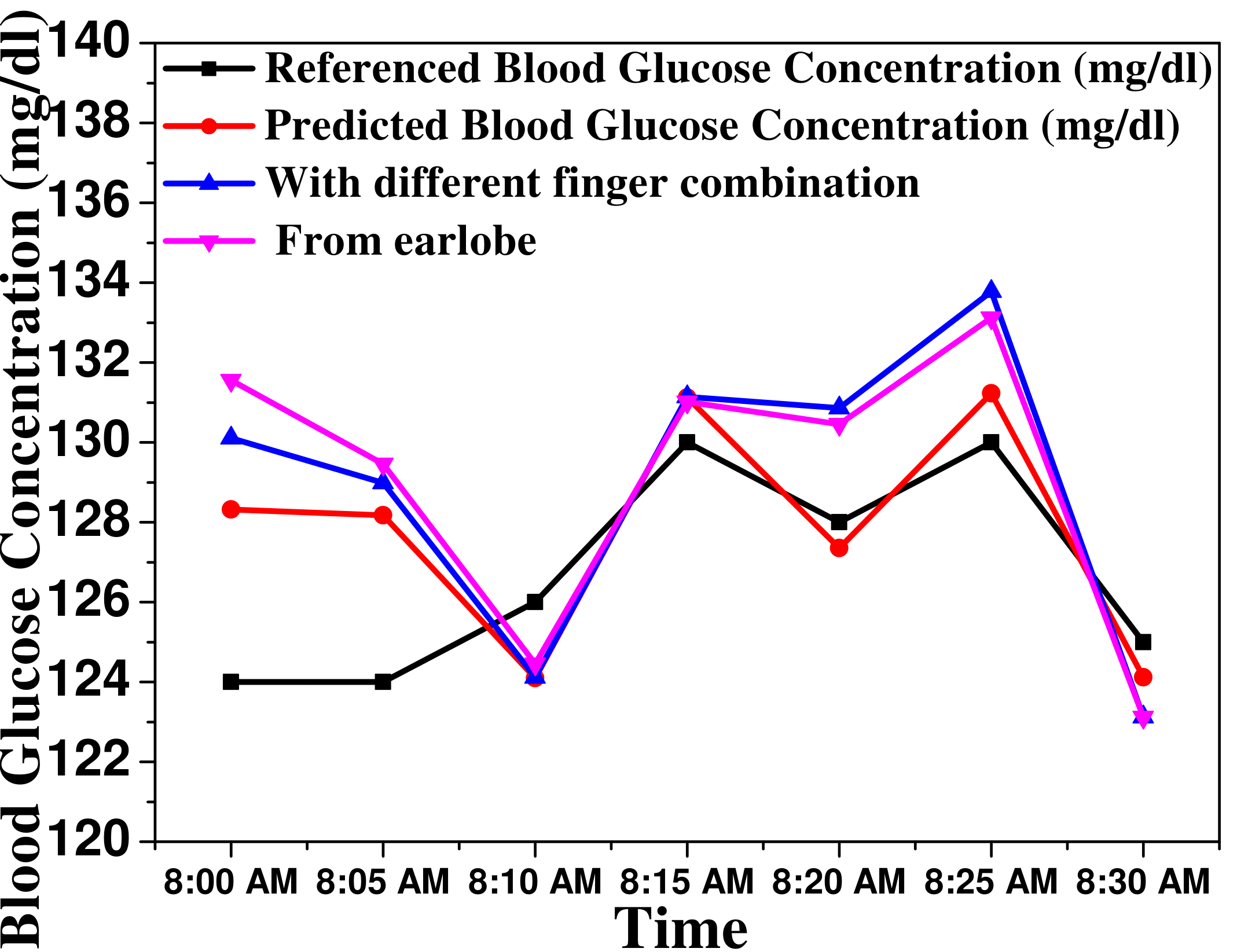}}\label{}
	\subfigure[Time interval of 60 minutes]{\includegraphics[width=0.44\textwidth]{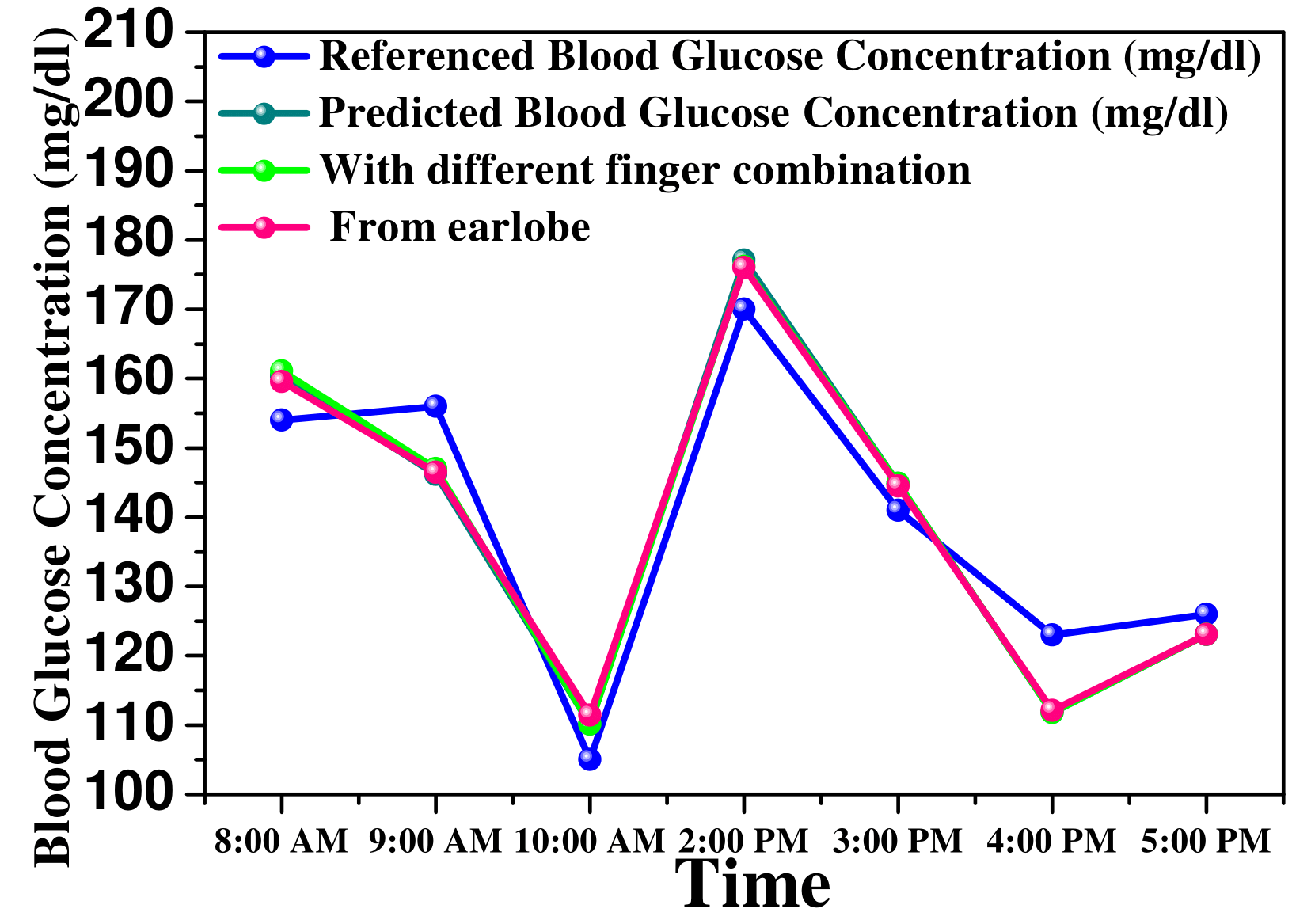}}\label{}
\caption{Predicted and reference blood glucose concentration for validation of iGLU on single volunteer.}
	\label{graph}
\end{figure*}

To test the device stability, experimental analysis has been done where multiple measurements of the same subject have been done a couple of times. For this experimental work, a volunteer has been recruited for measurement of blood glucose through iGLU and invasive method with time intervals of 5 minutes and the same experiment has been done for a time interval of two hours. The graphical representation is shown in Fig. \ref{graph}.

A volunteer has been taken for glucose measurement with a time interval of 5 minutes to check the stability of proposed non-invasive device iGLU. There was 10 mg/dl blood glucose deviation during 7 iterations of blood glucose measurement. During experimental analysis, it has been observed that it was 2-4 mg/dl blood glucose deviation between referenced and predicted blood glucose concentrations which is shown in Fig \ref{graph} (a). A different volunteer has also been taken for another experimental analysis to validate the non-invasive device accuracy which is shown in Fig.\ref{graph} (b). Glucose measurement has been done with a time interval of 60 minutes. Measurements have been done in 7 iterations. There are variations in reference blood glucose concentration between 8:00 AM-10:00 AM, 10:00 AM-2:00 PM and 2:00 PM-4:00 PM. These variations (low to high) represent the glucose intakes in the form of food. During this analysis, 5-10 mg/dl absolute deviation has been observed which represents the stability and precision level. During these experimental works, it has also been observed that the effect of objects (fingers or earlobes) changes is negligible. Hence, this device has been found stable and free from the constraint of objects for measurement.

\begin{figure*}[htbp]
	\centering
	\subfigure[Healthy]{\includegraphics[width=0.40\textwidth]{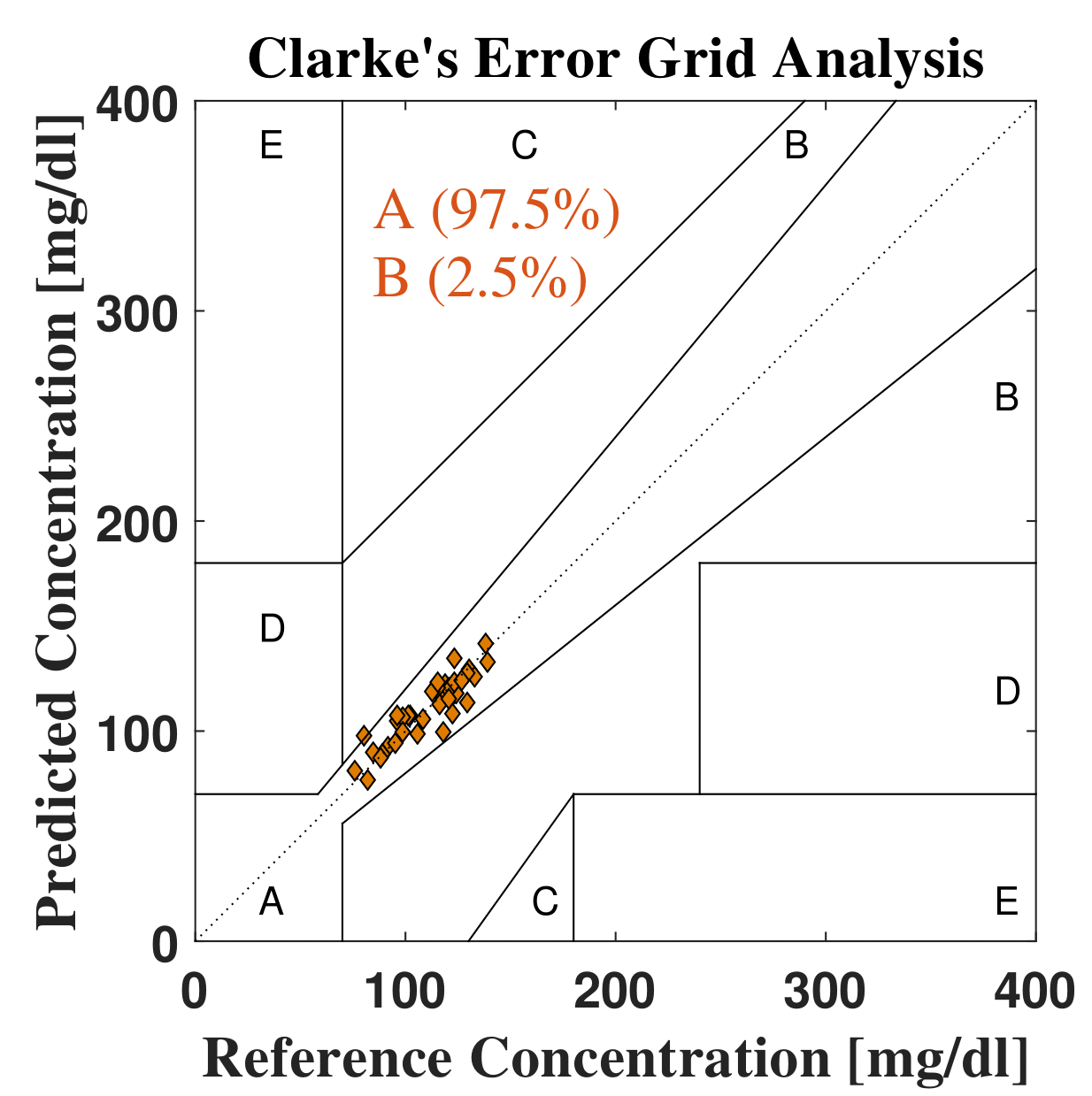}}\label{}
	\subfigure[Prediabetic]{\includegraphics[width=0.40\textwidth]{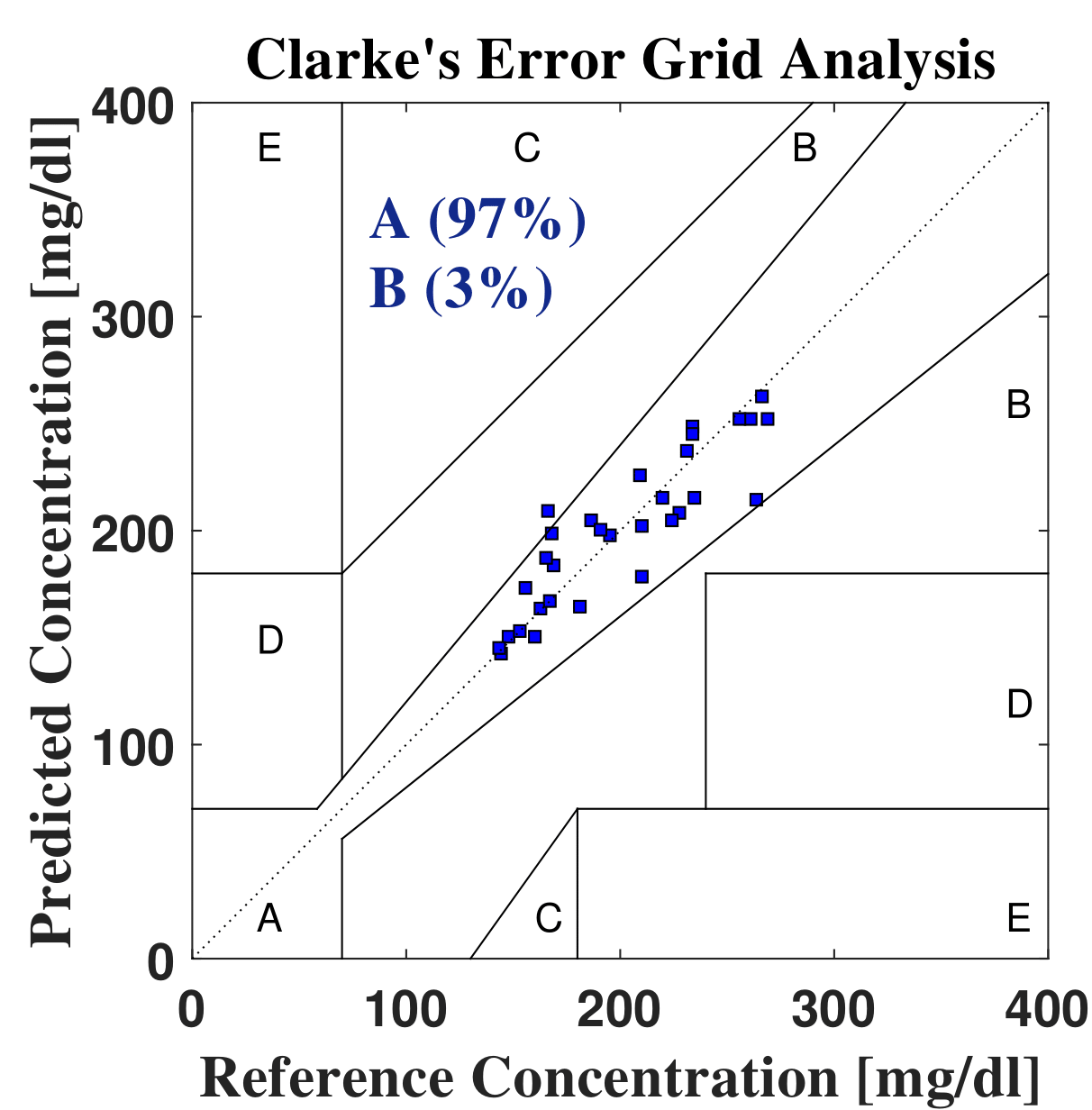}}\label{}\\
	\subfigure[Diabetic]{\includegraphics[width=0.40\textwidth]{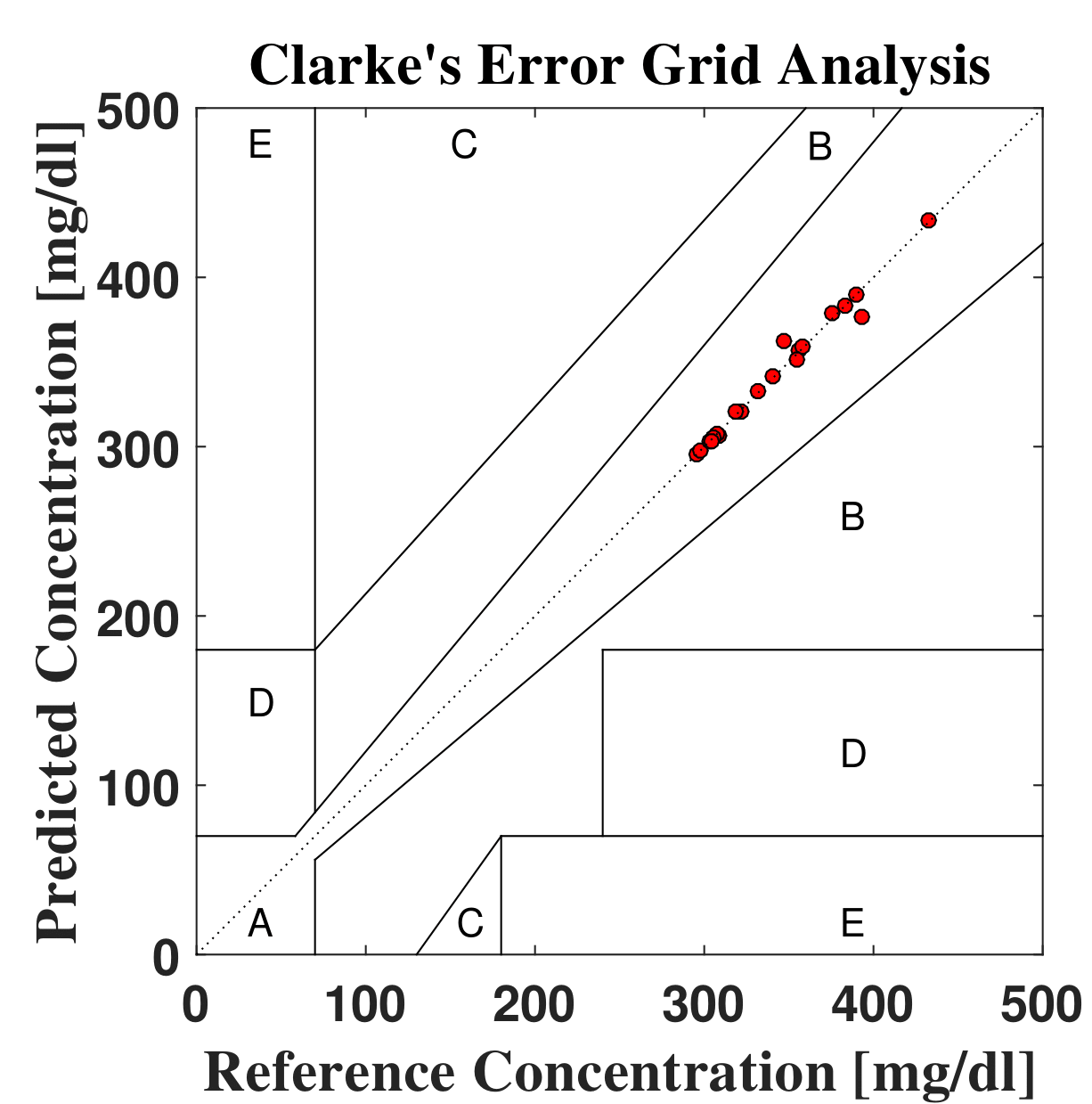}}\label{}
	\subfigure[Male samples]{\includegraphics[width=0.40\textwidth]{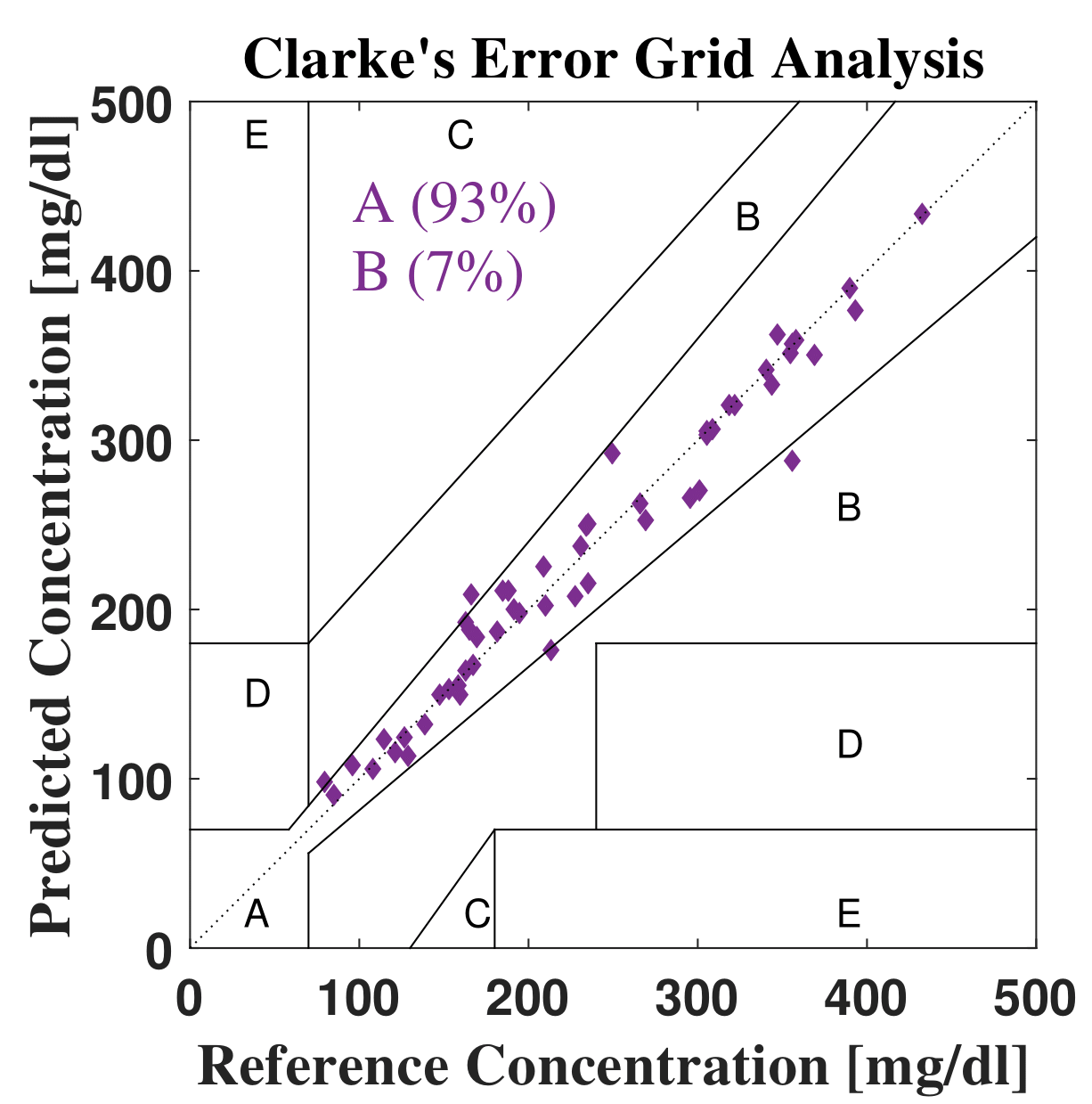}}\label{}\\
\subfigure[Female samples]{\includegraphics[width=0.40\textwidth]{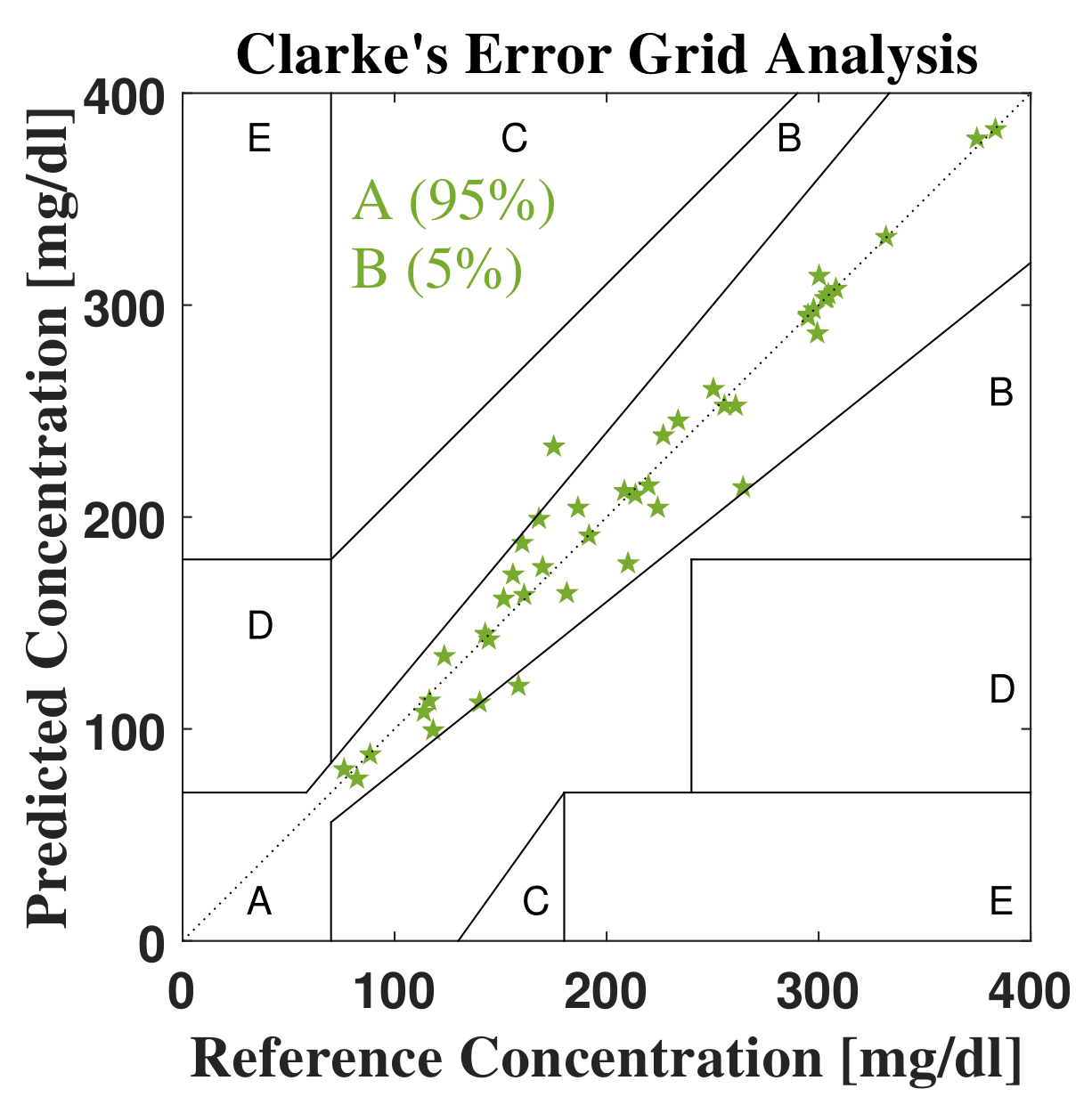}}\label{}
\subfigure[Validation from all samples]{\includegraphics[width=0.40\textwidth]{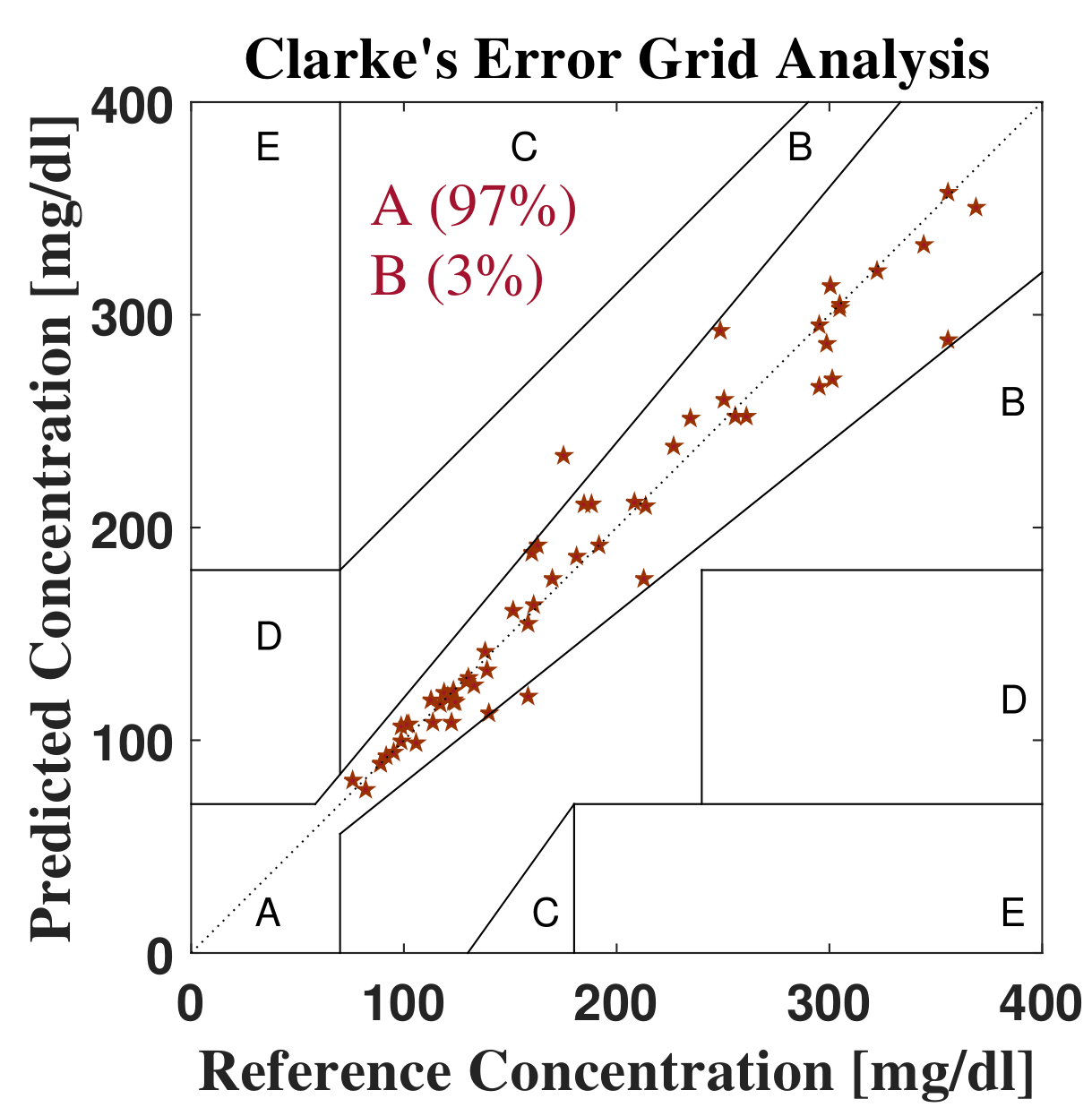}}\label{}
\caption{CEG analysis of predicted blood glucose concentration for validation of iGLU device using proposed calibrated MPR3 model.} 
	\label{CEG}
\end{figure*}

\begin{figure}[htbp]
	\centering
	\includegraphics[width=0.4\textwidth]{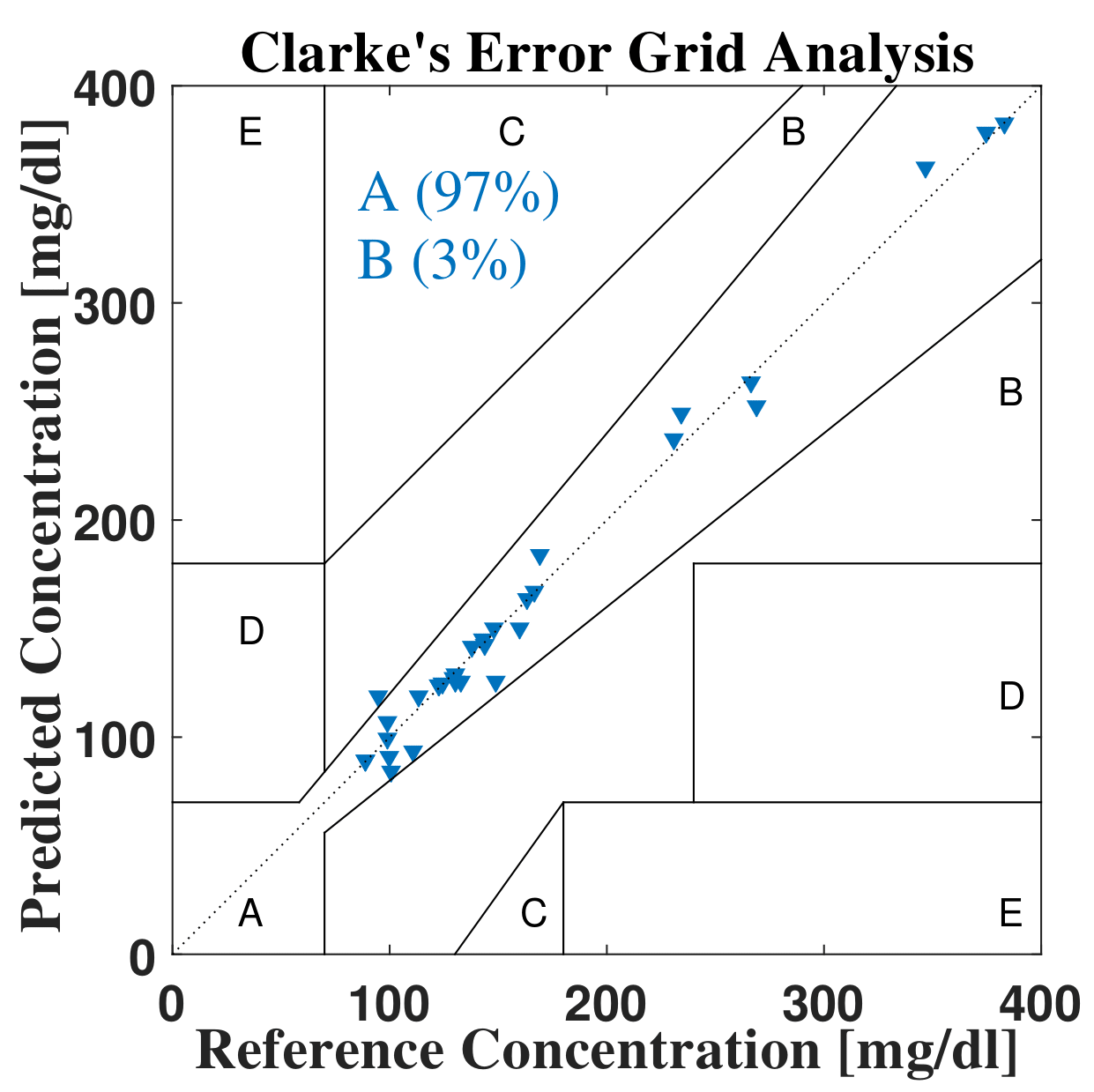}
\caption{CEG analysis of predicted blood glucose concentration for testing from independent samples using proposed calibrated MPR3 model.}
	\label{testing}
\end{figure}

Clarke Error Grid (CEG) analysis which is considered as a standard to categorize the biomedical devices has been used to analyze the accuracy of predicted blood glucose concentration values from proposed device \cite{Clarke2005}. Clarke analysis elaborates the zones by the difference between referenced and predicted blood glucose concentration. The predicted values enter in zone A and B; then the device will be desirable. In this analysis, samples have been arranged gender and glucose level category wise to check the accuracy of iGLU. Validation and testing have been done using the proposed calibrated MPR3 model.
After analysis of the proposed device, it has been found that all predicted values of glucose concentration exist in zone A and B. The proposed device is desirable for clinical purpose. The Clarke error grid analysis of the proposed device are represented in Fig. \ref{CEG} and \ref{testing}.

\begin{table}[htbp]
	\caption{Statistical Analysis of validation of proposed combination and existing techniques}
	\label{val}
	\centering
	\begin{tabular}{ccccc}
		\hline
		\textbf{}&$mARD$&$AvgE$&$MAD$&$RMSE$\\
		\textbf{}&\%&\%&mg/dl&mg/dl\\
		\hline		\hline
		Logistic&25.57&27.54&48.29&71.08\\
		\hline
		SVR&8.22&9.35&16.81&29.94\\
		\hline
		DNN&7.32&7.03&9.89&11.56\\
		\hline
		\textbf{MPR3}&\textbf{6.01}&\textbf{6.08}&\textbf{11.30}&\textbf{18.29}\\
		\hline
	\end{tabular}
\end{table}

According to Table \ref{table_example}, it is concluded that 4.66\% $mARD$ is better than the previous work. The device comprises of emitters, detectors, ADC and passive components such as resistances and capacitances which would be resulted in low cost (20-25 USD approx.) solution. Analysis and device validation have been done using human blood for real-time applications. In this proposed work, regression coefficient, average error, mean absolute deviation and root mean square error are also compared with related works. The average error and mean absolute deviation 4.61\% and 7.55 mg/dl show the accuracy of the proposed device.

\begin{table}[!h]
	\caption{Comparison with Previous Work}
	\label{table_example}
	\centering
	\begin{tabular}{ccccccccc}
		\hline
		\textbf{Works}&Regression&\textbf{$mARD$}&\textbf{$AvgE$}&\textbf{$MAD$}&$RMSE$&Category&CEG&Device\\
		\textbf{}& Coefficient&(\%)&(\%)&(mg/dl)&(mg/dl)& &(A\&B)&cost\\
		\textbf{}&($R^{2}$)& & &  & & & & \\
		\hline		 \hline
		Song et al.	\cite{Song2015}&-&8.3&19&-&-&Non&100\%&Cheaper\\
		& &&&&&invasive&&\\
		\hline
		Demitri et al. \cite{demitri2017measuring}&-&-&-&7.8 mg/dl&-&Invasive&100\%&-\\
		& &&&&&&&\\
		\hline
		Acciaroli et al.	\cite{acciaroli2018reduction}&-&8.94&-&-&-&Semi &97.6\%&Costly\\
		& &&&&&invasive&&\\
		\hline
		Wang et al. \cite{7933990}&0.97&-&-&-&-&Semi&-&-\\
		& &&&&&invasive&&\\
		\hline
		Gani et al.	\cite{5291722}&-&-&-&-&4.0&Semi&99\%&-\\
		& &&&&&invasive&&\\
		\hline
		Chee et al.	\cite{1186525}&-&-&-&$>$50.0&-&Semi&94.4\%&-\\
		& &&&&&invasive&&\\
		\hline
		\textbf{Proposed Work}&\textbf{0.81}&\textbf{4.66}&\textbf{4.61}&\textbf{7.55}&\textbf{11.95}&\textbf{Non}&\textbf{100\%}&\textbf{Cheaper}\\
		& &&&&&\textbf{invasive}&&\\
		\hline
	\end{tabular}
\end{table}


\section{Conclusions and Future Directions}

Multiple short-wave spectroscopy techniques based non-invasive blood glucose monitoring device is proposed and validated for real-life application. The combination of absorption and reflection of light at specific wavelengths using MPR based calibration is optimized for non-invasive measurement. The proposed device is an innovative approach in the form of a combination of spectroscopy techniques for precise detection. During statistical analysis using the proposed computation device, 0.81 correlation coefficient is calculated using MPR based calibration with polynomial degree 3. The $AvgE$ and $mARD$ using the proposed calibration method are improved as compared to other non-invasive glucose measurement device. After analysis of predicted blood glucose values, 100\% samples come in the zone A and B. During statistical analysis of possible combinations of techniques with proposed combination, it is concluded that the device with the combination of three channels is more optimized as compared to other possible combinations and precision measurement has been observed between the range of 70-450 mg/dl. 

In future research, we will add more features of Internet-of-Medical-Things (IoMT) integration of the iGLU. For example, integration of stress measurement along with blood-glucose level. Integration of control methods of stress and blood-glucose along with monitoring will be considered. Closed feedback from healthcare providers-end to the individual end-user side for management or control of effects when needed to ensure remote healthcare when there may be a shortage of healthcare providers. 
In future research, we will consider Internet-of-Medical-Things (IoMT) integration of the iGLU so that the data can be stored in the cloud. We will also evaluate the options of edge-computing and IoT-computing paradigms for the fast response as well as long-term storage of the user's records.  A combined blood glucose level monitoring along with automatic insulin injection through a pump can have a single unified solution that can positively affect the lives of millions around the planet. In such situation security of the devices for various reasons including wrong manipulation of insulin dosage and draining the wearable's battery are concerns which need research attention \cite{Rachakonda_ICCE_2019}. Overall the ongoing and future research can have a significant impact on the smart living component of smart healthcare for smart homes \cite{Sundaravadivel_CEM_2018-Jan}.

\section*{Acknowledgment}

The authors would like to thank Dispensary and System Level Design and Calibration Testing Lab, Malaviya National of Technology, Jaipur, Rajasthan (India) and special thanks for MHRD funded SMDP Lab which has provided the support of hardware components for the experimental implementation. We would also thank Dr. Navneet Agrawal (endocrinologist) and his team for the help and support at Diabetes, Obesity and Thyroid Centre, Gwalior (M.P.). Data collection has been done under inspection of him following medical protocols.

This article is an extended version of our article \cite{Jain_IEEE-MCE_2020-Jan_iGLU1}.

\bibliographystyle{IEEEtran}



\section*{Authors' Biographies}


\begin{minipage}[htbp]{\columnwidth}
\begin{wrapfigure}{l}{1.0in}
	\vspace{-0.4cm}
\includegraphics[width=1.0in,keepaspectratio]{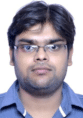}
	\vspace{-0.3cm}
\end{wrapfigure}
\noindent
\textbf{Prateek Jain} (GS'18) earned his B.E. degree in Electronics Engineering from Jiwaji University, India in 2010 and Master degree from ITM University Gwalior. He is a Research Scholar at th ECE department of MNIT, Jaipur. His current research interest includes VLSI design, Biomedical Systems and Instrumentation. He is an author of 14 peer-reviewed publications. He is a regular reviewer of 12 journals and 10 conferences.
\end{minipage}

\vspace{2.0cm}

\begin{minipage}[htbp]{\columnwidth}
\begin{wrapfigure}{l}{1.0in}
\vspace{-0.4cm}
\includegraphics[width=1.0in,keepaspectratio]{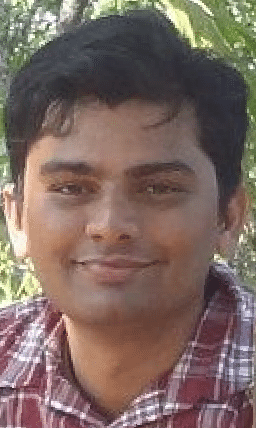}
	\vspace{-0.5cm}
\end{wrapfigure}
\noindent
\textbf{Amit M. Joshi} (M'08) has completed his M.Tech (by research) in 2009 and obtained Doctoral of Philosophy degree (Ph.D) from National Institute of Technology, Surat in August,2015. He is currently working as Assistant Professor at National Institute of Technology, Jaipur since July,2013. His area of specialization is Biomedical signal processing, Smart healthcare, VLSI DSP Systems and embedded system design. He has published six book chapters and also published 50+ research articles in excellent peer reviewed international journals/conferences. He has worked as a reviewer of technical journals such as IEEE Transactions, Springer, Elsevier and also served as Technical Programme Committee member for IEEE conferences. He also received UGC Travel fellowship , SERB DST Travel grant  and CSIR Travel fellowship to attend IEEE Conferences in VLSI and Embedded System. He has served session chair at various IEEE Conferences like TENCON -2016, iSES-2018, ICCIC-14 etc. He has already supervised 18 M.Tech projects and 14 B.Tech projects in the field of VLSI and Embedded Systems and VLSI DSP systems. Presently, He is supervising four full-time PhD students and two part-time PhD.
\end{minipage}

\vspace{2.0cm}
\begin{minipage}[htbp]{\columnwidth}
	\begin{wrapfigure}{l}{1.00in}
		\vspace{-0.3cm}
		\includegraphics[width=1.0in,keepaspectratio]{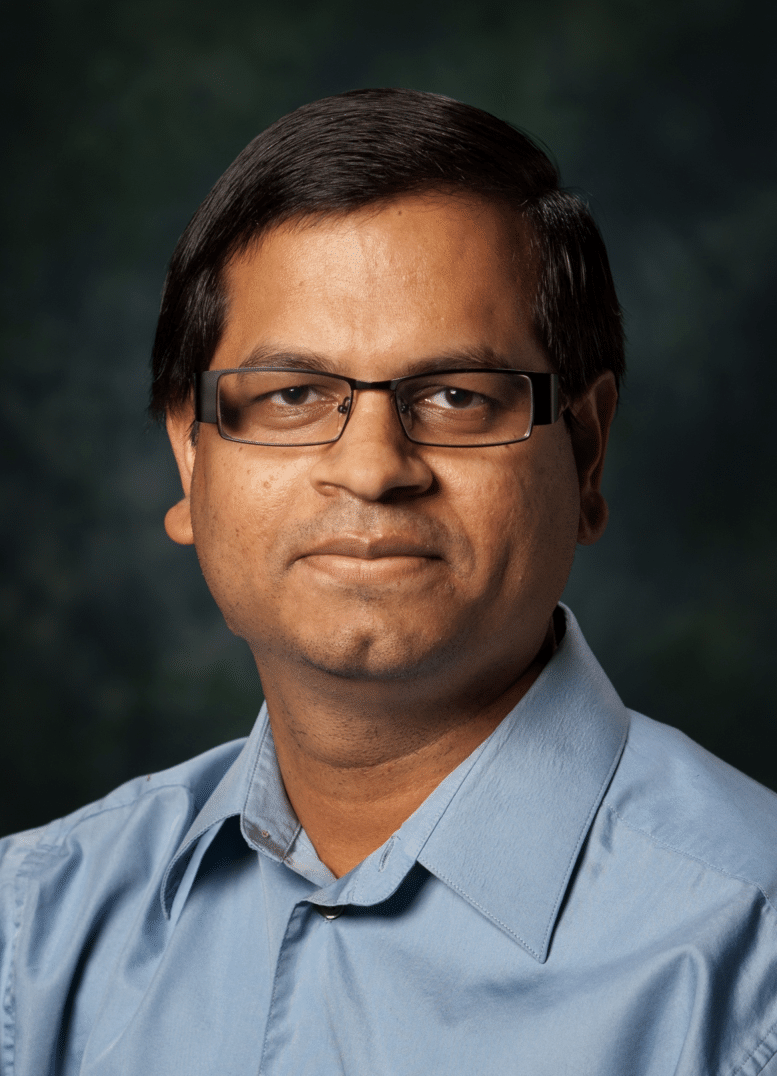}
		\vspace{-0.5cm}
	\end{wrapfigure}
\noindent
\textbf{Saraju P. Mohanty} (SM'08) received the bachelor's degree (Honors) in electrical engineering from the Orissa University of Agriculture and Technology, Bhubaneswar, in 1995, the master's degree in Systems Science and Automation from the Indian Institute of Science, Bengaluru, in 1999, and the Ph.D. degree in Computer Science and Engineering from the University of South Florida, Tampa, in 2003.
He is a Professor with the University of North Texas. His research is in ``Smart Electronic Systems'' which has been funded by National Science Foundations (NSF), Semiconductor Research Corporation (SRC), U.S. Air Force, IUSSTF, and Mission Innovation Global Alliance. He has authored 300 research articles, 4 books, and invented 4 U.S. patents. His has Google Scholar citations with an H-index of 32 and i10-index of 120. He was a recipient of ten best paper awards, IEEE Consumer Electronics Society Outstanding Service Award in 2020 for contributions to the IEEE Consumer Electronics society, the IEEE-CS-TCVLSI Distinguished Leadership Award in 2018 for services to the IEEE and to the VLSI research community, and the 2016 PROSE Award for Best Textbook in Physical Sciences and Mathematics category from the Association of American Publishers for his Mixed-Signal System Design book published by McGraw-Hill. 
He has delivered 9 keynotes and served on 5 panels at various International Conferences. He has been serving on the editorial board of several peer-reviewed international journals, including IEEE Transactions on Consumer Electronics (TCE), and IEEE Transactions on Big Data (TBD). 
He is currently the Editor-in-Chief (EiC) of the IEEE Consumer Electronics Magazine (MCE). 
He has been serving on the Board of Governors (BoG) of the IEEE Consumer Electronics Society, and has served as the Chair of Technical Committee on Very Large Scale Integration (TCVLSI), IEEE Computer Society (IEEE-CS) during 2014-2018. He is the founding steering committee chair for the IEEE International Symposium on Smart Electronic Systems (iSES), steering committee vice-chair of the IEEE-CS Symposium on VLSI (ISVLSI), and steering committee vice-chair of the OITS International Conference on Information Technology (ICIT). He has mentored 2 post-doctoral researchers, and supervised 10 Ph.D. dissertations, 26 M.S. theses, and 10 undergraduate projects. 
\end{minipage}

\end{document}